\documentclass[showpacs,amsmath,amssymb,twocolumn]{revtex4}
\usepackage{latexsym}
\usepackage{epsfig}
\usepackage{bm}
\usepackage{dcolumn}

\begin{document}

 \title{ Hadronization Approach for a Quark-Gluon Plasma Formed in Relativistic Heavy Ion Collisions }
 \author{Hong Miao$^1$, Chongshou Gao$^{2}$, Pengfei Zhuang$^1$}
 \affiliation{$^1$Physics Department, Tsinghua University, Beijing
100084, China} \affiliation{$^2$School of Physics, Peking
University, Beijing 100871, China}

\vspace{16pt}
\begin{abstract}
A transport model is developed to describe hadron emission from a
strongly coupled quark-gluon plasma formed in relativistic heavy
ion collisions. The quark-gluon plasma is controlled by ideal
hydrodynamics, and the hadron motion is characterized by a
transport equation with loss and gain terms. The two sets of
equations are coupled to each other, and the hadronization
hypersurface is determined by both the hydrodynamic evolution and
the hadron emission. The model is applied to calculate the
transverse momentum distributions of mesons and baryons, and most
of the results agree well with the experimental data at RHIC.

\end{abstract}

\maketitle

\section {Introduction} %
\label{sec:intro}
The hot and dense matter created in relativistic heavy ion
collisions provides the condition to investigate the new state of
matter of QCD, the so-called quark-gluon plasma(QGP)
\cite{sign:intro1,sign:intro2,sign:itro_Rafelski_phL,sign:EOS,sign:intro3}.
The space-time evolution of QGP and its hadronization process are
still open questions in the research of this area. Hydrodynamics
\cite{sign:hydro01,sign:hydro_Bro,sign:hydro_Nonaka} based on the
assumption of local thermal and chemical equilibrium, models of
quark coalescence or recombination \cite{sign:coal_1,sign:coal_2,
sign:reco_rev, sign:reco_1, sign:reco_2,sign:reco_S}, and
fragmentation \cite{sign:JetQuenchL1,sign:JetFrag1} are often used
to describe the evolution and hadronization of QGP formed in heavy
ion collisions at RHIC. However, we do not know a priori if the
colliding system can reach equilibrium or not. In fact, because of
the very small size and very short lifetime of the collision zone
\cite{sign:finitesize1,sign:exp_HBT_STAR_200pion,sign:hydro02},
the highly excited system may spend a considerable fraction of the
lifetime in a non-equilibrium state \cite{sign:Hdr_expand}. On the
other hand, even if the system reaches equilibrium in the early
quark-gluon stage, the hadrons formed in the later stage will be
in non-equilibrium state, since the particle density in hadron
phase is much smaller than that in the parton stage.

Evaporation models
\cite{sign:Eva_eva,sign:Eva_Cascade,sign:Eva_eqi} developed many
years ago are widely used to describe the hadronization of QGP. In
some of the models, the emission rate is calculated from a simple
assumption of balance between the particle emission and absorption
in QGP and in hadron gas at a critical temperature $T_c$. Under
this assumption, the particle emission distribution is just the
corresponding ideal thermal distribution of hadron gas, $ f({\bf
p}) = f^T({\bf p})$.

However, considering the feedback effect of hadron evaporation on
QGP, the space-time evolution of QGP and hadron evaporation are
coupled to each other and the system is governed by both the
hydrodynamic expansion and the evaporation. On the other hand,
when the expansion of the system is fast, the fluid will be more
likely to split into relatively small bubbles during the hadron
evaporation. In this case, the system is a mixture of QGP bubbles
and physical vacuum, and the energy density of the system is lower
than that of a unitary QGP. These bubbles may provide a mechanism
to reduce the viscosity of the system and make it more accessible
for the ideal hydrodynamics to work at
RHIC\cite{sign:viscosityexp}. Thirdly, from the analysis of
collective flow measured at RHIC \cite{sign:Idealfluid}, the QGP
formed at RHIC is a strongly coupled quark-gluon plasma (sQGP)
\cite{sign:diquark_baryon,sign:sign
bound_state,sign:CSC1,sign:DiquarkRevIn} where there exist not
only quarks and gluons but also some of their resonant states like
$qq$ and $qg$.

In this paper, we consider an effective hadronization model for
the production of soft particles in relativistic heavy ion
collisions. In the model, the QGP evolution is described by ideal
hydrodynamics together with the equation of state including
quarks, gluons and their resonant states. At the critical
temperature, the expansion does not change the temperature of the
system, and the decrease of the energy density is due to the
splitting of the QGP fluid. When the energy density reaches the
value for hadronization, which is a free parameter in our
calculation, the evaporation starts and the hadron distributions
are controlled by transport equations. The evaporation
hypersurface is characterized by both the hydrodynamics and the
evaporation.

The paper is organized as follows. In Section \ref{s2} we discuss
hadron production and transport and derive an approximate emission
distribution. In Section \ref{s3} we consider the hydrodynamics
for the evolution of QGP with possible splitting into droplets and
the effective evaporation boundary condition. We apply our model
to hadron spectra in heavy ion collisions at RHIC energy in
Section \ref{s4}. We summarize in Section \ref{s5}.

\section{Hadron Emission }
\label{s2}
We investigate in this section the transport of hadrons emitted
from QGP. The classical transport equation for the evolution of
hadron phase space distribution $f(t,{\bf r},{\bf p})$ can be
written as
\begin{equation}
 \frac{\partial f}{\partial t} + {\bf v}\cdot\nabla f =
 Q + \alpha - \beta f,
\end{equation}
where the second term on the left hand side is the free streaming
part, and on the right hand side the lose and gain terms
$\beta({\bf p}) f$ and $\alpha({\bf p})$ indicate hadron
absorption and production in QGP, and the elastic term $Q({\bf
p};f)$ \cite{sign:R_Boltzmann1} controls the thermalization of the
hadrons.

Under the assumption of quasi-static condition, the time evolution
\cite{sign:R_Boltzmann2time} during the transport can be
temporarily removed. If we only focus on a one dimensional hadron
transport in the region $x>0$ and treat the elastic scattering in
relaxation time approximation, the hadron transport can be greatly
simplified,
\begin{equation}
\label{trans}
 {v \cos \theta} {\partial f\over\partial x}  = -{f-f^T\over \tau} + \alpha - \beta f,
\end{equation}
where $v=p/E$ is the hadron velocity, $\tau$ the relaxation time
parameter, and $f^T(p)$ the thermal hadron distribution. In
principle, $\tau$ should depend on the hadron momentum and the
type of hadrons. For simplicity, we take it as a universal
constant in our global model. Considering the finite volumes of
hadrons and resonances, the phase space distribution will be
suppressed. An effective way to include the volume effect in the
thermal distribution is to multiply $f^T$ by a factor $1-y_o$,
where $y_o$ is the ratio of the occupation volume of all the
excluded hadrons to the whole volume of the system
\cite{sign:vol_lit, sign:VollEff}. In the following calculation we
will always consider this volume effect for $f^T$.

The transport equation (\ref{trans}) can be solved with the
analytic result
\begin{equation}
\label{analy} f(x,\theta,p) =
f_\infty(p)+\left[f(0,\theta,p)-f_\infty(p)\right]e^{-{\beta(p)\tau+1\over
\tau v\cos\theta}x}
\end{equation}
with $f_\infty(p)$ defined as
\begin{equation}
 f_\infty(p)={f^T(p)+\alpha(p)\tau\over \beta(p)\tau+1}.
\end{equation}
It is easy to see that the hadrons reach chemical equilibrium in
the limit of vanishing inelastic interactions with
$\alpha=\beta=0$ and thermal equilibrium in the limit of
infinitely strong elastic interaction with $\tau=0$. In the both
cases, the distribution at $x\to \infty$ is reduced to the thermal
one,
\begin{equation}
f_\infty(p) = f^T(p).
\end{equation}

Taking into account the boundary condition at $x=0$,
\begin{equation}
 f(0,\theta,p)= 0, \ \ \ \ (\cos\theta>0),
\end{equation}
which means no hadrons moving from the vacuum in the region of
$x<0$ to the QGP phase in the region of $x>0$, and the constraint
\begin{equation}
0\le f(x,\theta,p) \le f_\infty(p),
\end{equation}
the hadron distribution can be expressed as
\begin{equation}
 f(x,\theta,p)= f_\infty(p)\left[1-\Theta(\cos\theta)e^{-{\beta(p)\tau+1\over \tau v
 \cos\theta}x}\right],
\end{equation}
where $\Theta(x)$ is a step function.

It can be seen from the above discussion that the influence of the
transport process is mainly reflected in the region near the QGP
surface at $x=0$. Most of the emitted hadrons are created in this
region. Since the energy density in the inner part of the QGP
fluid is higher than the value on the surface, the hadronization
of the QGP fluid can be effectively regarded as a surface
evaporation.

In the above derivation we did not consider the QGP surface shift
and the breaking of the fluid in the process of hadron emission. A
simple way to take the surface shift and fluid splitting into
account is to introduce two parameters $v_s$ and $\eta$ in the
model. The former is the shrinking velocity of the surface, and
the latter is the ratio of the occupation volume of all the QGP
droplets to the whole volume of the system. These two parameters
should be considered simultaneously in the transport equation for
hadron emission and the hydrodynamic equation for QGP evolution.
With these two parameters, the hadron transport equation can be
effectively written as
\begin{equation}
\left(v\cos\theta-v_s\right){\partial f\over \partial x}
     = \eta\left[-{f-f^T\over \tau} + \alpha - \beta f\right]
\end{equation}
with the solution
\begin{eqnarray}
 f(x,\theta,p) &=& f_\infty(p)\times\\
 &&\left[1 - \Theta(v\cos\theta-v_s)e^{-{\beta(p)\tau+1\over
 \tau(v\cos\theta-v_s)}\int^x_0\eta(x')
 dx'}\right].
 \nonumber
\end{eqnarray}

With the hadron distribution, the energy loss of the QGP fluid in
the rest frame of the boundary can be expressed as
\begin{equation}
\label{eloss} {dE\over dtdS} = -\sum_i \int
f^i_\infty(p_i)\left(v_i\cos\theta-v_s\right)E_i d^3{\bf p}_i,
\end{equation}
where the integration over the angle $\theta$ is restricted by the
condition $v_i\cos\theta-v_s < 0$, $E_i$ is the energy of a hadron
of type $i$, and the summation runs over all the hadron types.
Similarly, the momentum loss of the QGP fluid can be written as
\begin{equation}
\label{ploss} {dP\over dtdS} = \sum_i \int
f^i_\infty(p_i)\left(v_i\cos\theta-v_s\right)p\cos\theta d^3{\bf
p}_i.
\end{equation}

If hydrodynamic expansion is neglected, the equation
\begin{equation}
\label{eloss2}
{dE\over dt dS} = \epsilon v_s
\end{equation}
defines a minimal shrinking velocity of the fluid, where
$\epsilon$ is the average energy density at the boundary. When
$\epsilon$ is small enough to make $v_s = 1$, the local
association between different parts of the fluid will completely
vanish, and the fluid must decouple. Therefore, the condition for
the fluid description of the system is that the energy density
must satisfy
\begin{equation}
\epsilon \ge \left.{dE\over dtdS}\right|_{v_s=1}.
\end{equation}

The question left for the emission distribution of hadrons is the
determination of the loss and gain terms $\beta$ and $\alpha$. We
first calculate the absorption rate of hadrons by the medium.
Considering that tens of kinds of hadrons will be discussed, the
calculation of their different absorptions would be quite
complicated. For simplicity, the inelastic absorption rate of a
hadron by partons of type $j$ is described as
\begin{eqnarray}
\beta_j({\bf p}) &=& \int g_j({\bf q})|{\bf v}-{\bf v}_q|\sigma_j({\bf p},{\bf q})d{\bf q}\\
&=& 4\pi \int q^2g_j(q)\sigma_j({\bf p},{\bf
q})\frac{(v+v_q)^3-|v-v_q|^3}{6vv_q}dq,\nonumber
\end{eqnarray}
where $v_q=q/E$ denotes the parton velocity, $g_j$ is the parton
thermal distribution at the critical temperature $T_c$ determined
by the hydrodynamic equation, and $\sigma_j({\bf p},{\bf q})$ is
the effective absorption cross section. Replacing $\sigma_j({\bf
p},{\bf q})$ by its average value $\sigma_j$, we obtain the total
absorption rate
\begin{equation}
 \beta(p) = 4\pi \sum_j{\sigma_j\zeta_j(p)}
\end{equation}
with
\begin{eqnarray}
\zeta_j(p) &=& \int_0^{\frac{m_jv}{\sqrt{1-v^2}}}
\left[vg_j(q)q^2 +{1\over 3v}g_j(q)\frac{q^4}{E_j^2}\right]dq\nonumber\\
&&+\int_{\frac{m_jv}{\sqrt{1-v^2}}}^\infty\left[g_j(q)\frac{q^3}{E_j}+\frac{v^2}{3}g_j(q)qE_j\right]dq.
\end{eqnarray}
Since the diquark dynamics is not yet very clear, to simplify the
calculation we take the assumption that the baryon cross section
is $1.5$ times the meson cross section,
$\langle\sigma_b\rangle=1.5 \langle\sigma_m\rangle$.

For high momentum hadrons passing cold matter, we have
\begin{eqnarray}
 \zeta_j(p) &\approx& v\int_0^{\infty}{g_j(q)q^2dq}
            + \frac{1}{3v}\int_0^{\infty}{g_j(q)\frac{q^4}{E_j^2}dq}\nonumber\\
            & = & \frac{n_j}{4\pi}\left[v + \frac{1}{3v}\langle{v_j^2}\rangle\right],
\end{eqnarray}
where, $\langle {v_j^2} \rangle$ is a constant at given
temperature and chemical potential, and $n_j$ is the parton
density. In this case, the total rate can be approximately
expressed as
\begin{equation}
\beta(p) \approx {v\over L},
\end{equation}
where $L = 1/\sum_i n_j\sigma_j$ is the inelastic mean free path.

For soft hadrons in hot medium, we have
\begin{eqnarray}
 \zeta_j(p) & \approx & \int_{0}^\infty{g_j(q)\frac{q^3}{E_j}dq}
            + \frac{v^2}{3}\int_{0}^\infty{g_j(q)qE_jdq}\nonumber\\
            & = &\frac{n_j}{4\pi}\left[\langle{v_j}\rangle+\frac{v^2}{3}\langle\frac{1}{v_j}\rangle\right].
\end{eqnarray}
Taking $\langle{v_j}\rangle = \langle\frac{1}{v_j}\rangle = 1$ for
gluons and light quarks and $\langle{v_j}\rangle \approx 0.96$ and
$ \langle\frac{1}{v_j}\rangle \approx 1.05$ for strange quarks
with mass $m_s = 95$ MeV, the total rate becomes approximately
\begin{equation}
\beta(p)\approx {1+v^2/3\over L}.
\end{equation}

There are many mechanisms to describe hadron production in heavy
ion collisions, such as two- or three-quark recombination for
mesons or baryons\cite{sign:reco_1, sign:reco_2,sign:reco_S}. In
our model, there are not only quarks but also diquarks in the QGP.
We assume that, mesons and intermediate baryons are respectively
made of two quarks and one quark and one diquark, and final state
baryons are considered as combinations of different intermediate
baryon states\cite{sign:diquark_baryon}, see Appendix
\ref{app:wavefunc}.

The produced hadron density in coordinate and time space through
combination of two constituents $j$ and $k$ at a fixed channel $I$
can be defined as \cite{sign:crossection}
\begin{equation}
n_I=\int|v_j-v_k|g_j({\bf p}_j)g_k({\bf p}_k)\sigma(p_j,p_k,
p)d^3{\bf p}_jd^3{\bf p}_k,
\end{equation}
where $g_j$ and $g_k$ are the constituent distributions. and
$\sigma$ is the cross section,
\begin{eqnarray}
 \label{eq:cross}
 \sigma(p_j,p_k,p)&=&(2\pi)^4\int\frac{|M_I|^2}{4\sqrt{(p_j\cdot p_k)^2-m_j^2m_k^2}}\\
 &&\times\delta^4(p_j+p_k-p)\delta(p^2-m^2)\Theta(E)\frac{d^4p}{(2\pi)^3}\nonumber
 \end{eqnarray}
with the  scattering matrix element $M_I$.

For baryon production, since a diquark is much heavy than the
corresponding quark, the hadron density can be approximately
expressed as \cite{sign:diquark_baryon,sign:diquark_Omega},
\begin{equation}
n_I=2\pi^{3}|M_I|^2 \int g_j(E_j)g_k(E_k)dE_jdE_k,
\end{equation}
where the integrated region is restricted by the constraints
\begin{eqnarray}
m_j &\leq& {E_j}\leq{\infty},\\
E_jE_k &\geq& \frac{(E_j+E_k)(m_j^2E_k+m_k^2E_j)}{m^2}+E_c^2,\nonumber\\
E_c &=& {1\over
2m}\sqrt{\left[m^2-(m_j+m_k)^2\right]\left[m^2-(m_j-m_k)^2\right]}.\nonumber
\end{eqnarray}

Since we have assumed thermalization of quarks, gluons and
diquarks, their distribution functions can be written as
\begin{equation}
g_j({\bf p}_j) = g_j(E_j)=
\frac{\kappa}{(2\pi)^3}\frac{\omega_j}{e^{\frac{E_j-\mu_j}{T_c}}\pm
1},
\end{equation}
where $\omega_j=3J_j(J_j+1)$ is the degeneracy factor of spin and
color degrees of freedom, and $\kappa$ is the reduction factor
resulted from the volume effect of the bound-state or
resonance-state components \cite{sign:VollEff},
\begin{equation}
\kappa =\left[(1-y_o)+C_t y_o\right]^{2-l}(1-y_o)^l,\nonumber\\
\end{equation}
with $l$ being the number of bound states or resonant states and
$C_t$ the "transparency" of bags. For simplicity, we set $C_t=1$.

Making summation over all the possible channels, we obtain the
final hadron density
\begin{equation}
\label{ci1}
n = \sum_I C_I n_I,
\end{equation}
where $C_I$ is the weight factor of channel $I$ and is listed in
Appendix \ref{app:wavefunc} for different channels.

To have detailed information on the produced hadrons, we need the
hadron momentum distribution
\begin{equation}
{dn_I\over 4\pi{p^2}dp} = {\pi^2\over 2}{|M_I|^2\over pE}\int
F(E_k)g_j(E-E_k)g_k(E_k)dE_k,
\end{equation}
where $F$ is defined as
\begin{equation}
F(E_j)={\sqrt{E E_k(m^2-m_j^2+m_k^2)-(E^2m_k^2+E_k^2m^2)}\over
mE_c},
\end{equation}
and the integrated region is restricted by the constraint
\begin{eqnarray}
E_{max}&\leq& E_k\leq E_{min},\nonumber\\
E_{max}&=&Max\{m_k,E_-\},\nonumber\\
E_{min}&=&Min\{E-m_j,E_+\},\nonumber\\
E_{\pm}&=&{E\left(m^2-m_j^2+m_k^2\right)\pm2mpE_c\over 2m^2}.
\end{eqnarray}

With the known produced hadron density, the gain term $\alpha$ in
the transport equation (\ref{trans}) can be written as
\begin{equation}
\label{ci2}
\alpha(p)=\sum_IC_I\frac{dn_I}{(4\pi{p^2})dp}.
\end{equation}

Note that in the calculation of lose and gain terms above, we
considered only hadronization and hadron interactions with the QGP
constituents. When the hadron decay in the transport process is
included, it will certainly change the loss and gain terms. Its
contribution to the production rate $\alpha$ will be effectively
discussed in Section \ref{s4}, and the change in the absorption
rate $\beta$ can be described by a shift \cite{sign:PDG2006}
\begin{equation}
\label{dloss}
\beta\to \beta+{\Gamma\over \gamma},
\end{equation}
where $\Gamma$ is the hadron decay width and $\gamma$ is the
Lorentz factor of the hadron.

In principle, all the particle masses are temperature and chemical
potential dependent \cite{sign:Latt_CrossOver,sign:TC1}. For
quarks and gluons at the hadronization, we can simply take
$m_g=0$, $m_u=3$ MeV, $m_d=6$ MeV and $m_s=95$ MeV
\cite{sign:PDG2006}, but for light hadrons, their masses are in
principle controlled by chiral symmetry and therefore deviate from
their vacuum values. To simplify calculations, we will still use
their masses and decay widths in the vacuum and neglect the medium
effect. As for the diquark and $qg$ states, their masses in the
vacuum \cite{sign:VollEff} are taken as $m_{qq}=485$ MeV,
$m_{qs}=580$ MeV in $0^+$ channel, $m_{qq}=725$ MeV, $m_{qs}=820$
MeV, $m_{ss}=905$ MeV in $1^+$ channel, and $m_{qg}=610$ MeV,
$m_{sg}=715$ MeV in ${1\over 2}^-$ channel. These values are a bit
smaller than the estimation in constituent quark model
\cite{sign:Diquarkmassdef}.

\section{ QGP Evolution }
 \label{s3}
We now turn to the space-time evolution of QGP. For ideal QGP, it
can be described by the hydrodynamic equations,
\begin{equation}
\label{hydro}
\partial^\mu T_{\mu\nu}=0,\ \ \ \ \partial^\mu j_\mu^i=0,
\end{equation}
where $T_{\mu\nu}=(\epsilon+P)u_\mu u_\nu-Pg_{\mu\nu}$ is the
energy-momentum tensor of the fluid, $j^i_\mu=n_iu_\mu$ refers to
the conserved charge current like baryon current and strange
current, and $\epsilon, P, n_i$ and $u_\mu$ are respectively the
energy density, pressure, conserved charge density and four
velocity of the fluid.

To close the hydrodynamic equations, we need the equation of state
of the fluid, $\epsilon(P)$. Since the parton fluid formed at RHIC
energy is a strongly coupled QGP, we include in the equation of
state the $u, d, s$ quarks, gluons, and all possible $qq$ and $qg$
resonant states. To simplify the calculation, we assume the
relation $\epsilon = 3 P$ in the QGP stage.

From the hydrodynamic equation (\ref{hydro}) and the equation of
state, we can obtain the energy density $\epsilon$ as a function
of space and time. When $\epsilon$ reaches $\epsilon_c$
corresponding to the critical temperature $T_c$ for deconfinement,
the expansion of the system will no longer change the temperature,
and the decrease of $\epsilon$ is assumed due to the splitting of
the QGP fluid. Therefore, for the expansion process at fixed
temperature $T_c$, the ratio of the energy densities
$\epsilon/\epsilon_c$ is always less than $1$. Finally, when the
energy density drops down to the value $\epsilon_b$ for the
hadronization, which is a free parameter in our calculation, the
hadron emission starts and the hydrodynamic evolution of the
system becomes
\begin{equation}
\partial^\mu\left(T_{\mu\nu}-\overline{T}_{\mu\nu}\right)=0,\ \ \ \
\partial^\mu\left(j^i_\mu-\overline{j}^i_\mu\right)=0,
\end{equation}
where $\overline T_{\mu\nu}$ and $\overline j^i_\mu$ are the
changes in energy-momentum tensor and conserved charge current due
to the hadron emission, and they are determined by the energy loss
(\ref{eloss}) and momentum loss (\ref{ploss}). In this way, the
evolution of the decoupling hypersurface of the fluid is
determined by both the hydrodynamics and the evaporation, similar
to the method in FCT calculations
\cite{sign:hydro_cal1,sign:hydro_cal2}.

We discuss now, as an example, the spherical expansion of
hydrodynamics to demonstrate the influence of evaporation on the
hadronization boundary. The $1+1$ hydrodynamics of QGP can be
written as \cite{sign:hydro_t_3},
\begin{eqnarray}
\label{hr}
 {\partial \xi\over \partial t}&=& -\frac{2}{3-v^2}\left[v{\partial \xi\over\partial r}+2{\partial v\over\partial r}+2(N-1){v\over r}\right],\nonumber\\
 {\partial v\over\partial t}&=&-{1\over 3-v^2}\bigg[{3(1-v^2)^2\over 4}{\partial\xi\over\partial r}+2v{\partial v\over\partial r}\nonumber\\
&&\ \ \ \ \ \ \ \ -(N-1){v^2(1-v^2)\over r}\bigg],\nonumber\\
 {\partial n_\gamma\over \partial t} &=& -\left[v{\partial
 n_\gamma\over\partial r}+n_\gamma{\partial v\over \partial r}+(N-1)n_\gamma{v\over
 r}\right]
\end{eqnarray}
for the spherical hydrodynamic velocity $v$, the scaled
energy-momentum quantity $\xi=\ln\left[(\epsilon + P)/(\epsilon_b
+ P_b)\right]$, and the conserved charge density
$n_\gamma=n_B\gamma$, where $\gamma$ is the Lorentz factor of the
fluid, $n_B$ is the baryon density, and $N=3$ is the symmetric
dimension.

Similarly, we can derive the simplified 1+1 hydrodynamic equation
including the feedback from the hadron evaporation. When
$\epsilon$ calculated from the hydrodynamics with evaporation is
less than $\epsilon_b$, the hadronization hypersurface is
determined by (\ref{hr}). Otherwise, the hydrodynamics with
evaporation itself gives the hypersurface.

\begin{figure}[h]
\centering
\includegraphics[width=.4\textwidth]{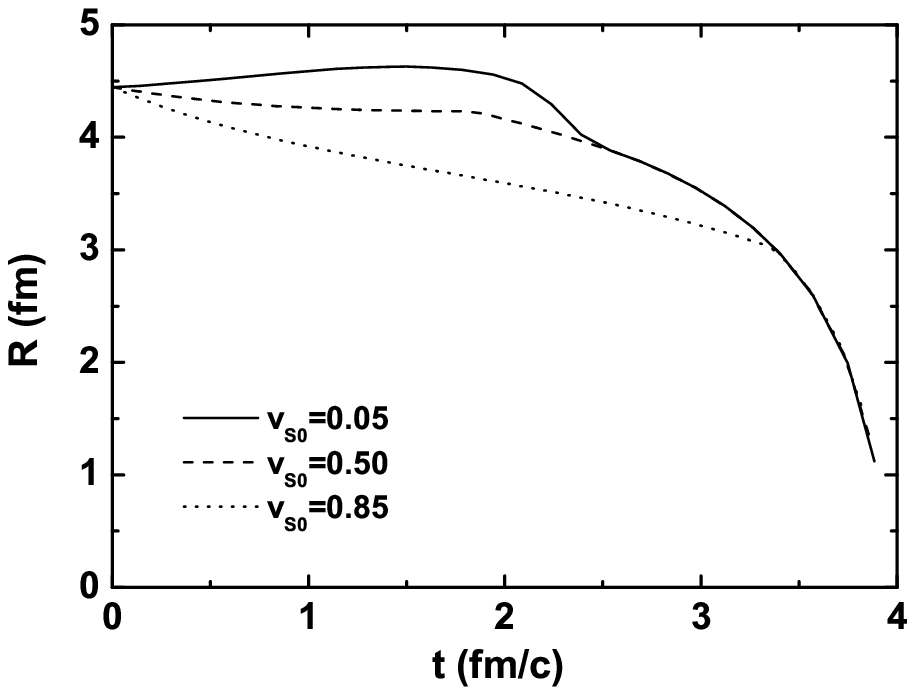}
\includegraphics[width=.4\textwidth]{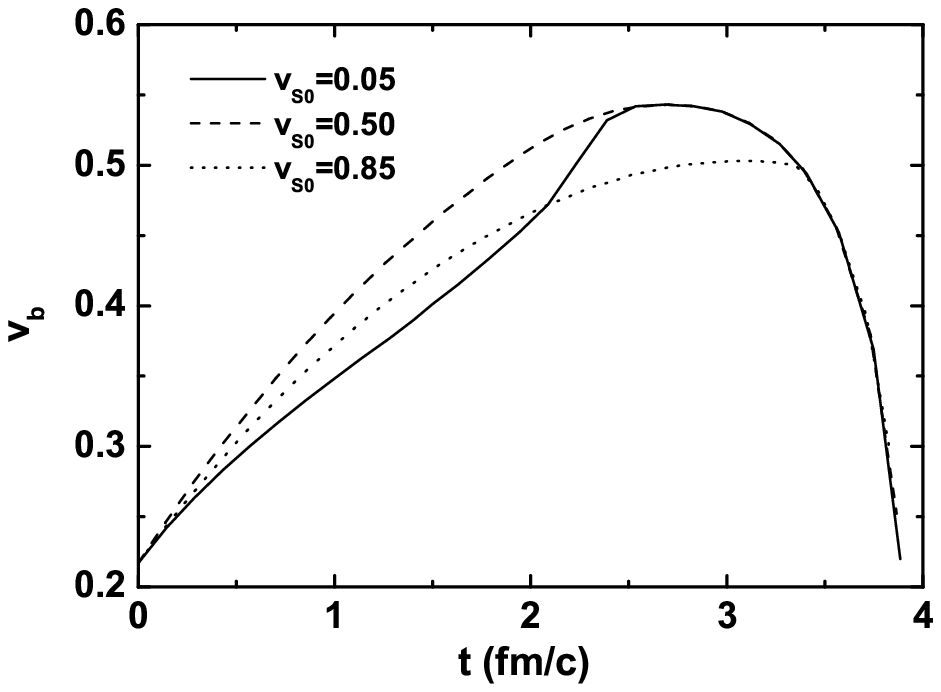}
\includegraphics[width=.4\textwidth]{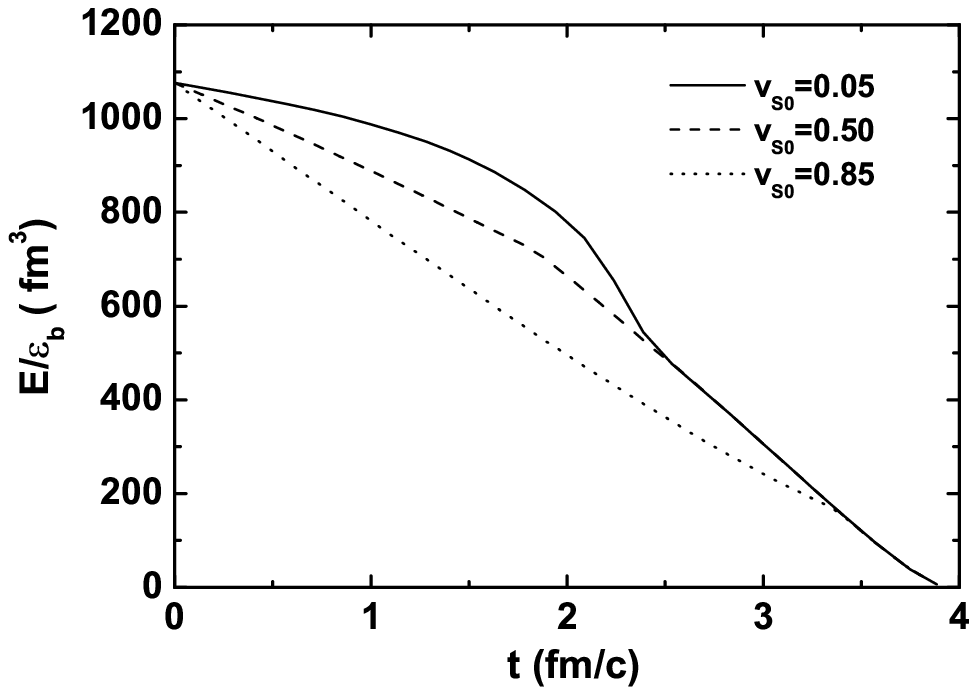}
\caption{The hadronization radius $R$, the expansion velocity $v$
at the radius, and the bulk energy $E$ scaled by the hadronization
energy density $\epsilon_b$ of the QGP fluid as functions of time
for several values of shrinking velocity $v_s$, calculated in the
frame of spherical expansion.} \label{fig1}
\end{figure}

The radius $R$ of the hadronization spherical surface, the
hydrodynamic expansion velocity $v$ on the surface, and the bulk
energy of the fluid are shown as functions of time in
Fig.\ref{fig1} for different values of shrinking velocity $v_s$
which characterizes the effect of hadron emission on the shift of
hydronization hypersurface. In the calculation, we have chosen the
initial energy density distribution as a Gaussian distribution
with the central value $\epsilon(r=0)=9\epsilon_b$ and the initial
expansion velocity $u(r)=\gamma v=0.05 r\ fm^{-1}$. For $v_s=0$
which means no hadron emission, the radius and the expansion
velocity are fully determined by the hydrodynamic equations
(\ref{hr}). With increasing hadron emission, the shrinking
velocity increases, and the hadronization radius, the expansion
velocity and the bulk energy deviate from the pure hydrodynamic
solution more and more strongly. Since the hydrodynamic expansion
grows with increasing time, the quantities are always controlled
by the pure hydrodynamics and the emission effect can be neglected
in the later stage of the fluid.

To study the space-time evolution of relativistic heavy ion
collisions at RHIC energy, it is better to consider the
cylindrical expansion of hydrodynamics. As is usually done, the
longitudinal expansion can be decoupled from the transverse
expansion and described in the Bjorken boost invariant scenario
\cite{sign:bjorken}. The transverse expansion is described by
equation (\ref{hr}) with $N=2$. As an example, we consider a
central Au-Au collision at colliding energy $\sqrt s =200$ GeV.
The initial condition of the collision, which is the same as in
\cite{sign:hydro01,sign:hydro02}, is determined by the nuclear
geometry \cite{sign:hydro_Glauber} and the corresponding
nucleon-nucleon interaction. The formation time of the locally
thermalized fluid is set to be $\tau_0 = 0.6$ $fm/c$.

\begin{figure}[h]
\centering
\includegraphics[width=.4\textwidth]{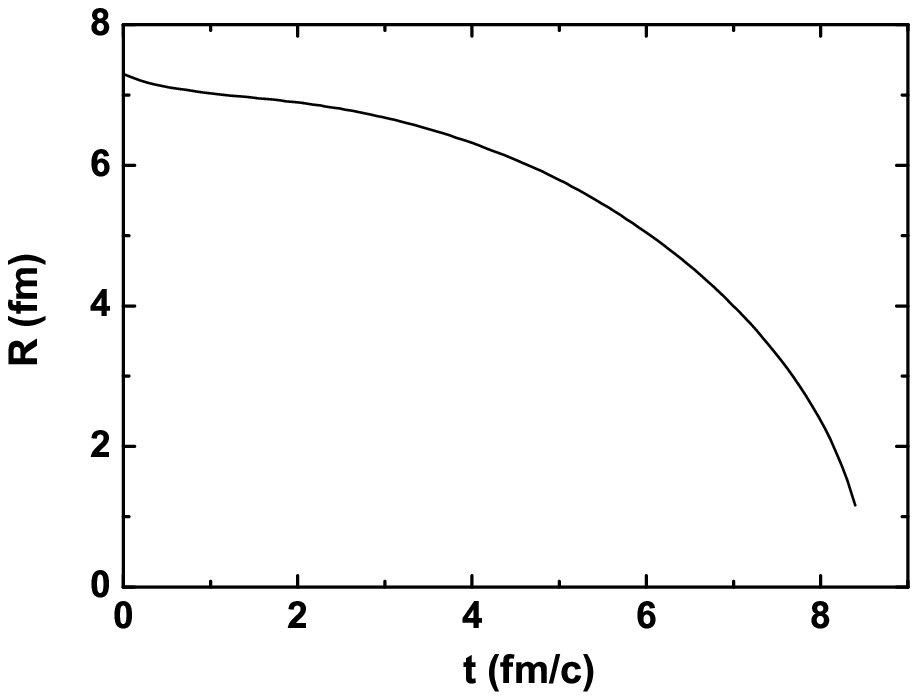}
\includegraphics[width=.4\textwidth]{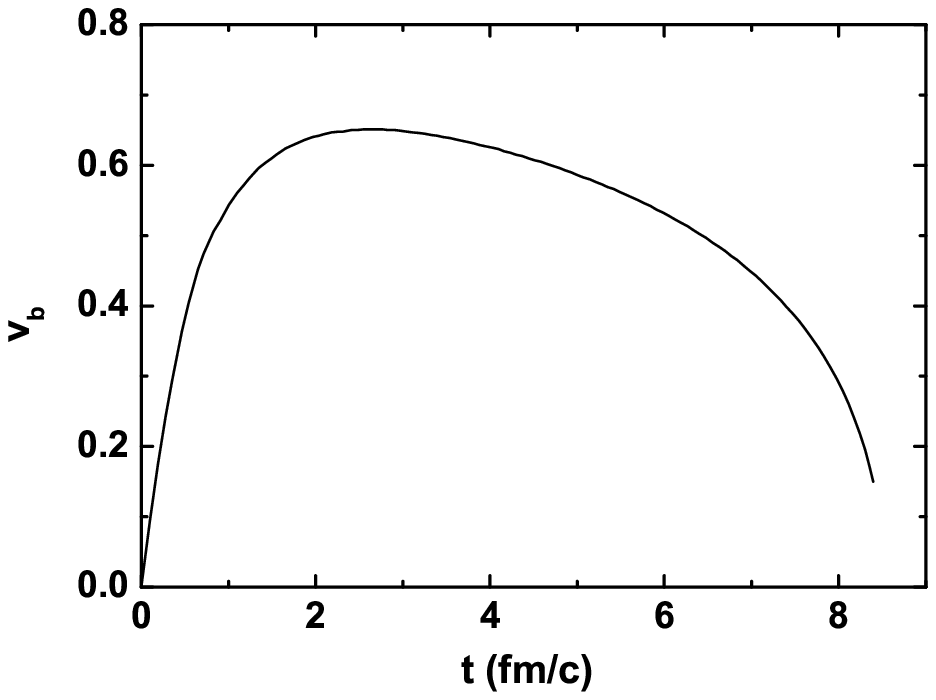}
\caption{The hadronization radius $R$ and the corresponding
expansion velocity $v$ of the QGP fluid as functions of time in a
central Au-Au collision with colliding energy $\sqrt s =200$ GeV
per pair of nucleons, calculated in the frame of cylindrical
expansion.} \label{fig2}
\end{figure}

The hadronization radius $R$ in the transverse plane and the
corresponding expansion velocity $v$ at $R$ are shown in
Fig.\ref{fig2}. The hadronization energy density is chosen as
$\epsilon_b=0.21$ GeV/fm$^3$ by fitting the RHIC data, and the
shrinking velocity $v_s$ is the maximum of its value solved from
the self-consistent equations (\ref{eloss}) and (\ref{eloss2}) for
the emission energy and its value determined from the QGP
hydrodynamic equation with splitting. Different from the spherical
expansion where the hadron emission plays an essential role in
determining the hadronization radius, the evaporation in the frame
of cylindrical expansion does not change the pure hydrodynamic
result considerably, except in the very beginning period. This
behavior is due to the rapid expansion in the Bjorken scenario.

\section{Hadron Production at RHIC}
 \label{s4}
Now it is the time for us to apply our model with hydrodynamics
for QGP evolution and transport for hadron emission to
relativistic heavy ion collisions. We will calculate the spectra
of soft hadrons produced in central Au-Au collisions at colliding
energy $\sqrt{s_{NN}}=200$ GeV and compare the results with
experimental data at RHIC \cite{sign:exp_PHENIX_ppik,
sign:exp_PHOBOS_pi, sign:exp_STAR_ppik2006,
sign:exp_STAR_StrangeOld, sign:exp_STAR_Strange,
sign:exp_STAR_Strange2, sign:exp_STAR_K892, sign:exp_STAR_phi,
sign:exp_PHENIX_phi, sign:exp_STAR_Sig1385, sign:exp_STAR_Xi1530,
sign:exp_Ratio_Omegaphi, sign:exp_Ratio_STAR_Omega}.

Let's first list the free parameters in the model. We choose the
critical temperature $T_c=166$ MeV which leads to the critical
energy density $\epsilon_c=1.91$ GeV/fm$^3$. Since resonant states
of $qq$ and $qg$ are taken into account in our equation of state
for the strongly coupled QGP, the energy density is larger than
the Stefan-Boltzmann limit for the ideal system of quarks and
gluons, $\epsilon_c=1.24\epsilon_{SB}$. The average cross section
for meson-quark inelastic scattering is taken to be
$\langle\sigma_m\rangle=7$ mb. The ratio $y_o$ to describe the
finite volume effect of hadrons is taken as $0.83$. From our
numerical calculation, the obtained hadron spectra are not
sensitive to the value of $y_o$ in a wide region. To minimize the
free parameters, we assume that the hadrons are not thermalized at
all, namely we take the relaxation time $\tau=\infty$ for all
hadrons. In this case, the hadron distribution is controlled only
by the emission rate and medium absorption,
\begin{equation}
 f_\infty(p)={\alpha(p)\over \beta(p)}.
\end{equation}
Further more, by fitting the PHENIX data
\cite{sign:exp_PHENIX_ppik} of protons and pions, we determine the
energy density of the QGP fluid $\epsilon_b=0.21$ GeV/fm$^3$ at
hadronization which corresponds to the ratio
$\eta=\epsilon_b/\epsilon_c=0.11$, the chemical potential
$\mu_B=27$ MeV at hadronization, and the meson-quark and
baryon-quark coupling constants $g_M^2=5.11$ and $g_B^2=34.24$.
For strange quark $s$, its chemical potential is
$\mu_s=\mu_B/3-\mu_S$ where $\mu_S$ is the strange chemical
potential. With the above parameters, the evolution duration of
the QGP fluid, calculated from the hydrodynamic equation, is about
$8.4$ fm/c.

The invariant yields of hadrons with type $i$ can be calculated
from the phase space integration of the hadron distribution
\begin{eqnarray}
\label{pty} \frac{d^2N_i}{2\pi p_Tdp_Tdy}
 &=&\frac{d}{2\pi p_Tdp_T}\sum_j \int f^j_\infty({\bf
 q}')\left(v'_j\cos\theta-v'_s\right)\nonumber\\
 &&\times  F_D^{ji}({\bf q},{\bf p})d^2{\bf x}dt'd^3{\bf q'},
\end{eqnarray}
where $F_D^{ji}$ stands for the decay from hadrons of type $j$ to
the hadrons of type $i$, ${\bf q}'$ is the hadron momentum in the
local rest frame of the hadronization hypersurface and is related
to the momentum ${\bf q}$ by a Lorentz transformation. It is
necessary to note that, the decay contribution to the gain term
$\alpha$, mentioned in the last section, can be effectively
described by multiplying the decay term $F_D^{ji}$ here by a
factor
\begin{equation}
\alpha_{ji}=1+\xi_{j+1,j}+\xi_{j+1,j}\xi_{j+2,j+1}+\cdots+\xi_{j+1,j}\cdots\xi_{i,i-1},
\label{eq:ITDgain}
\end{equation}
with
\begin{equation}
\xi_{j+1,j} = {\Gamma_j/\left(\gamma_j\eta\right)\over
\beta_{j+1}+\Gamma_{j+1}/\left(\gamma_{j+1}\eta\right)},
\end{equation}
where we have considered the assumption of $\tau=\infty$.

The phase space integration is done by a Monte-Carlo simulation,
which partially trace the directly emitted hadrons. By counting
enough events, the statistical error for $dN/dy$ caused by the
random simulation is less than $0.5\%-1\%$ for different hadrons.
The statistical error for $p_T$ distribution is, however,
relatively larger, especially in very low $p_T$ and high $p_T$
regions. The reason is the denominator $p_T$ in (\ref{pty}) for
low $p_T$ hadrons and low abundance for high $p_T$ hadrons. In the
region of $p_T=0.3-4$ GeV, the statistical error is expected to be
smaller than $5\%$ for most hadrons.

The light hadron yield is dominated by decay product. For example,
the number of direct pions is less than $5\%$ of the total final
state pion abundance. Since the convergence of heavy resonance
decay contribution is slow, many resonances need to be considered
in our calculations. The selection of hadron types is subject to
their relative production rates and branch ratios, compared with
their daughter particles counted in our calculations. Mesons and
baryons included in our model are listed in Table \ref{tab:meson}
and \ref{tab:baryon}. Although some of $0^+$ particles, such as
$f_0(600)$, $f_0(980)$ and $a_0(980)$ are often interpreted as
non-$q\bar{q}$ states, they are still treated as scalar mesons in
our calculations.
\begin{table}[h]
    \begin{center}
      \begin{tabular}{c c c c c}
        \hline\hline $0^-$ & $1^-$ & $0^+$ & $1^+$ & $2^+$ \\
        \hline
                     $\pi$   & $\rho$        & $a_{0\;980}$  & $b_{1\;1235}$ & $K_{2\;1430}$\\
                     $K$     & $K^*_{\;892}$ & $f_{0\;600}$  & $h_{1\;1170}$ &\\
                     $\eta$  & $\omega$      & $f_{0\;980}$  & $a_{1\;1260}$ &\\
                     $\eta'$ & $\phi$        & $K_{0\;1430}$ & $f_{1\;1285}$ &\\
                             & $K^*_{\;1410}$&               & $K_{1\;1270}$ &\\
                             &               &               & $K_{1\;1400}$ &\\
        \hline\hline
      \end{tabular}
    \end{center}
 \caption{Mesons selected in the calculations.}
 \label{tab:meson}
\end{table}
\begin{table}[h]
    \begin{center}
      \begin{tabular}{c c c c c c}
        \hline\hline $N$ & $\Delta$ & $\Lambda$ & $\Sigma$ & $\Xi$ & $\Omega$\\
        \hline
                     $N_{938}$  & $\Delta_{1232}$ & $\Lambda_{1116}$ & $\Sigma_{1192}$ & $\Xi_{1317}$ & $\Omega_{1672}$\\
                     $N_{1440}$ & $\Delta_{1600}$ & $\Lambda_{1405}$ & $\Sigma_{1385}$ & $\Xi_{1530}$ &\\
                     $N_{1520}$ & $\Delta_{1620}$ & $\Lambda_{1520}$ & $\Sigma_{1660}$ & $\Xi_{1690}$ &\\
                     $N_{1535}$ &                 & $\Lambda_{1600}$ & $\Sigma_{1670}$ & $\Xi_{1820}$ &\\
                                &                 & $\Lambda_{1670}$ & $\Sigma_{1750}$ & &\\
                                &                 & $\Lambda_{1690}$ & $\Sigma_{1775}$ & &\\
                                &                 & $\Lambda_{1830}$ &                 & &\\
        \hline\hline
      \end{tabular}
    \end{center}
 \caption{Baryons selected in the calculations.}
 \label{tab:baryon}
\end{table}

\begin{figure}[h]
    \centering
    \includegraphics[width=0.45\textwidth]{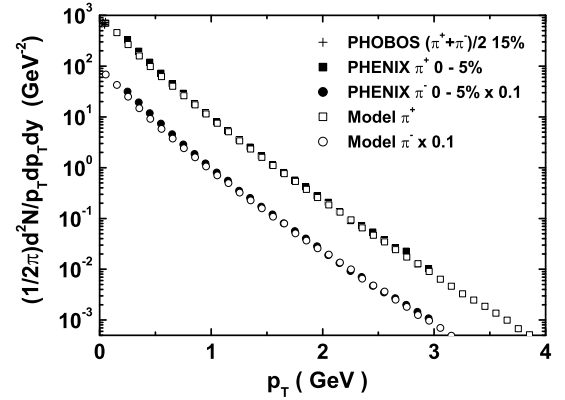}
    \caption{The transverse momentum distributions of $\pi^{\pm}$.
    The abundances are inclusive, and no feed-down corrections are applied.
    The data are from PHENIX \cite{sign:exp_PHENIX_ppik}
    and PHOBOS \cite{sign:exp_PHOBOS_pi} Collaborations. }
    \label{fig3}
\end{figure}

\begin{figure}[h]
    \centering
    \includegraphics[width=0.45\textwidth]{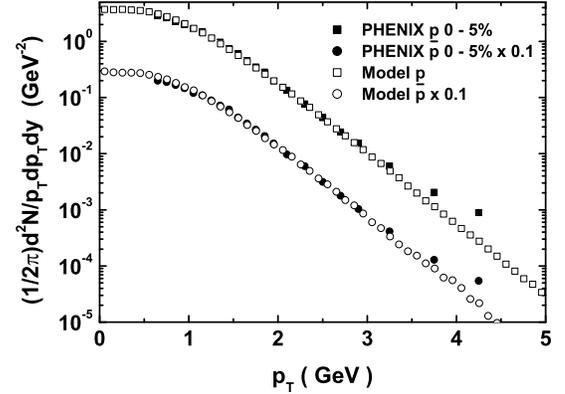}
    \caption{The transverse momentum distributions of protons and anti-protons.
    The feed-down corrections from $\Lambda$ are applied. The data are from PHENIX Collaboration \cite{sign:exp_PHENIX_ppik}.}
    \label{fig4}
\end{figure}

\begin{figure}[h]
    \centering
    \includegraphics[width=0.45\textwidth]{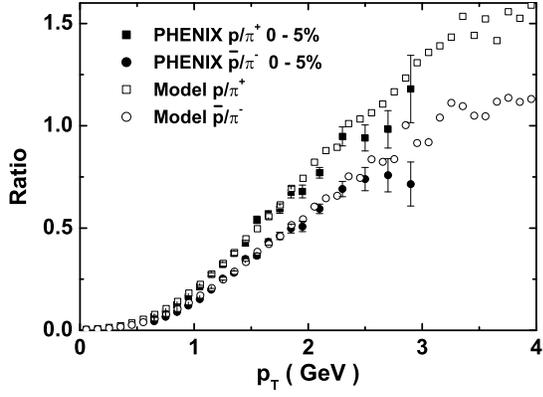}
    \caption{The $p/\pi^+$ and $\bar{p}/\pi^-$ ratios as functions of transverse momentum.
    The data are from PHENIX Collaboration \cite{sign:exp_PHENIX_ppik}. }
    \label{fig5}
\end{figure}

The hadron spectra and relative ratios are calculated with the
invariant momentum distribution (\ref{pty}) at $y=0$. Since the
data for exclusive protons and inclusive pions from PHENIX
\cite{sign:exp_PHENIX_ppik} are used to fit the parameters, the
agreement between our model calculation and the data for these two
kinds of particles, shown in Figs.\ref{fig3}, \ref{fig4} and
\ref{fig5}, looks perfect.

\begin{figure}[h]
    \centering
    \includegraphics[width=0.45\textwidth]{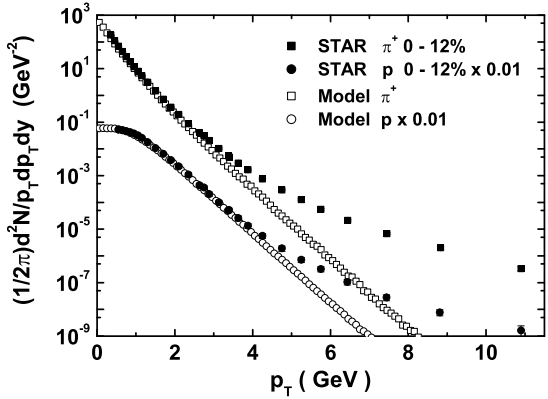}
    \caption{The transverse momentum distributions of proton and $\pi^+$. The distribution
    for proton is inclusive, and the decay contribution
    from $\Lambda$ and $K^0_S$ to pions is removed.
    The data are from STAR Collaboration \cite{sign:exp_STAR_ppik2006}. }
    \label{fig6}
\end{figure}

\begin{figure}[h]
    \centering
    \includegraphics[width=0.45\textwidth]{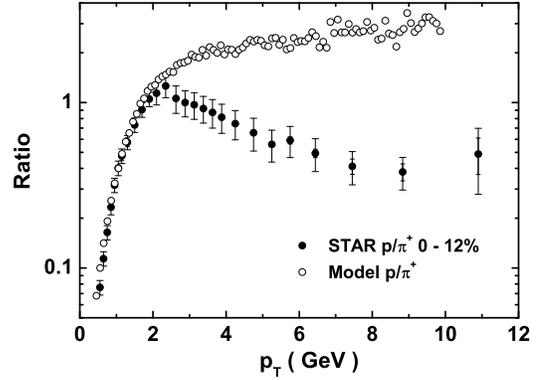}
    \caption{The $p/\pi^+$ ratio as a function of transverse
    momentum. The distribution
    for proton is inclusive, and the decay contribution
    from $\Lambda$ and $K^0_S$ to pions is removed.
    The data are from STAR Collaboration \cite{sign:exp_STAR_ppik2006}.}
    \label{fig7}
\end{figure}

As only soft particles are emitted from the QGP fluid itself, the
hadron spectra in high transverse momentum region are expected not
to be reproduced correctly in our model. For these high $p_T$
particles, we should consider the jet motion in QGP and its
fragmentation mechanism. The effect of jets or shower partons in
high $p_T$ region is more significant for pions and protons,
compared with those heavy resonances, especially multi-strange
heavy resonances. The big difference between our model calculation
and STAR data \cite{sign:exp_STAR_ppik2006} for high $p_T$ pions
and protons is shown clearly in Figs. \ref{fig6} and \ref{fig7}.

\begin{figure}[h]
    \centering
    \includegraphics[width=0.45\textwidth]{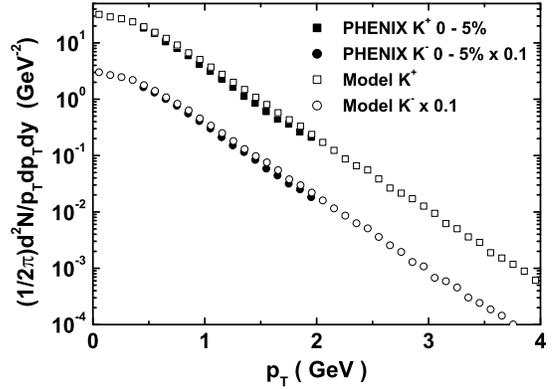}
    \caption{The transverse momentum distributions of $K^{\pm}$.
    The abundances are inclusive and no feed-down corrections are applied. The data are from
    PHENIX Collaboration \cite{sign:exp_PHENIX_ppik}.}
    \label{fig8}
 \end{figure}

The transverse momentum distributions for charged kaons are shown
in Fig. \ref{fig8}. Our model result looks higher than the
experimental data. The reason is probably the lack of heavy
resonances in our hadron list or the light strange quark with $m_s
= 95$ MeV \cite{sign:PDG2006}. A heavier strange quark can reduce
the production of strange hadrons like kaons, $\Sigma^*(1385)$ and
$\Xi^*(1530)$. For $m_s = 170$ MeV, the suppression for kaons is
about $5\%$. Another possible reason for the overestimation of
kaons is the strange hadron thermalization. If we take the
relaxation time $\tau\sim1$ fm, the model calculation agrees well
with the data.

\begin{figure}[h]
    \centering
    \includegraphics[width=0.45\textwidth]{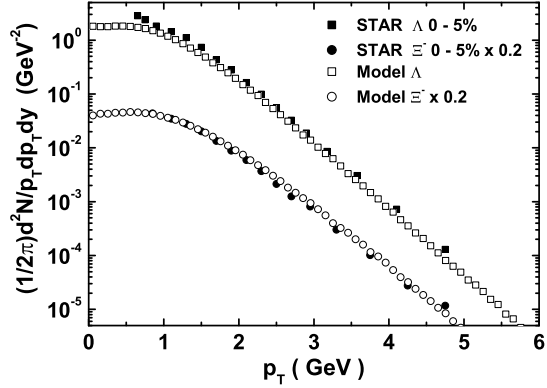}
    \caption{The transverse momentum distributions of $\Lambda$ and $\Xi^-$.
    The decay contribution from $\Xi^-$, $\Xi^0$ and $\Omega$ to $\Lambda$ is removed,
    but the one from $\Sigma^0$ is included. The distribution for $\Xi^-$ is inclusive. All the data are from
    STAR Collaboration \cite{sign:exp_STAR_StrangeOld,sign:exp_STAR_Strange}.}
    \label{fig9}
\end{figure}

\begin{figure}[h]
    \centering
    \includegraphics[width=0.45\textwidth]{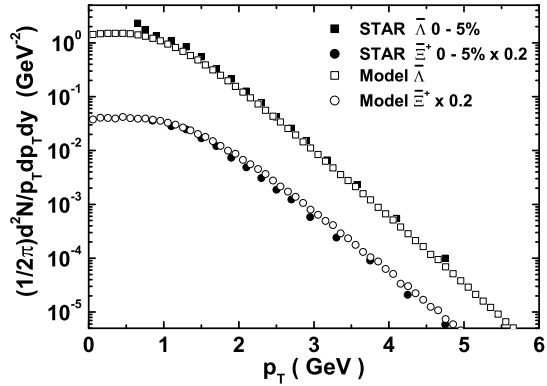}
    \caption{The transverse momentum distributions of $\bar{\Lambda}$ and $\bar{\Xi}^+$.
     The treatment for decay contribution is the same as in Fig.\ref{fig9}. The data are from
     STAR Collaboration \cite{sign:exp_STAR_StrangeOld,sign:exp_STAR_Strange}.}
    \label{fig10}
\end{figure}

The $\Lambda$ and $\Xi^-$ spectra are shown in Figs. \ref{fig9}
and \ref{fig10}. The model describes well the $\Xi$ data, but the
result deviates from the data for $\Lambda$ and $\bar{\Lambda}$
remarkably in low $p_T$ region, especially for the positive
$\Lambda$. Compared with the experimental data, a general behavior
of our model calculation is the enhancement of strange mesons and
suppression of strange baryons. This is probably due to the
exclusion of the hadron rescattering, such as
$N+\bar{K}\leftrightarrows\Sigma(\Lambda)+\pi$ or
$N+\bar{K}\rightarrow\Sigma^*(\Lambda^*)+X$, or the treatment of
diquarks. In our current frame, diquarks are thermalized like
quarks and gluons and satisfy the Bose-Einstein distribution.
Since diquarks are so heavy, it is more reasonable to describe
their phase-space distribution by a transport equation, like the
one for hadrons.

\begin{figure}[h]
    \centering
    \includegraphics[width=0.45\textwidth]{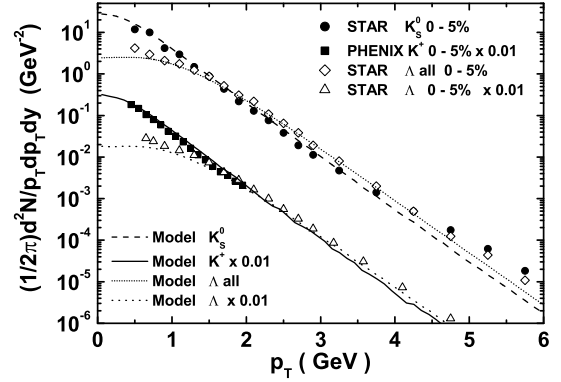}
    \caption{The transverse momentum distributions of ${\Lambda}, K^+$ and
    $K^0_S$. The distributions for $K^0_S$ and "$\Lambda$ $all$" are inclusive, and
    the other decay treatment is the same as in Fig.\ref{fig9}. All the data are from
    STAR Collaboration
    \cite{sign:exp_PHENIX_ppik,sign:exp_STAR_StrangeOld,sign:exp_STAR_Strange,sign:exp_STAR_Strange2}.}
    \label{fig11}
\end{figure}

\begin{figure}[h]
    \centering
    \includegraphics[width=0.45\textwidth]{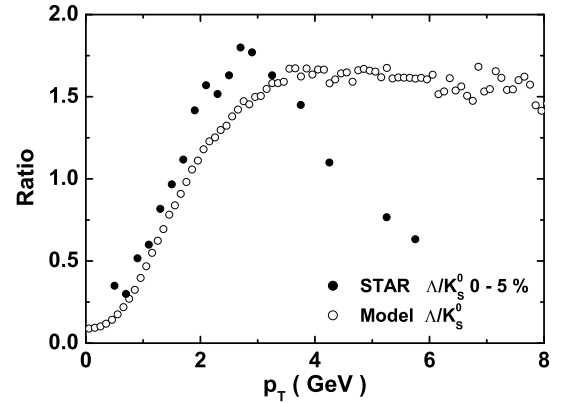}
    \caption{ The ratio $\Lambda/K^0_S$ as a function of transverse momentum.
    The abundances are inclusive, and no feed-down correction is applied. The data are from
    STAR Collaboration \cite{sign:exp_STAR_Strange2}. }
    \label{fig12}
\end{figure}

The transverse momentum distributions for $\Lambda, K^+$ and
$K^0_S$ and the ratio $\Lambda/K_S^0$ are shown in Figs.
\ref{fig11} and \ref{fig12}. Again we see clearly the theoretical
enhancement for $K$ and suppression for $\Lambda$. From the
comparison with the STAR data, the inclusive $\Lambda$
distribution without the feed-down correction seems much better
than the one shown in Fig.\ref{fig9}. Different from the
theoretical soft $p/\pi$ ratio shown in Fig.\ref{fig7} where the
ratio never decreases, the soft $\Lambda/K$ ratio begins to drop
down at about $p_T\sim 4$ GeV.

\begin{figure}[h]
 \centering
    \includegraphics[width=0.45\textwidth]{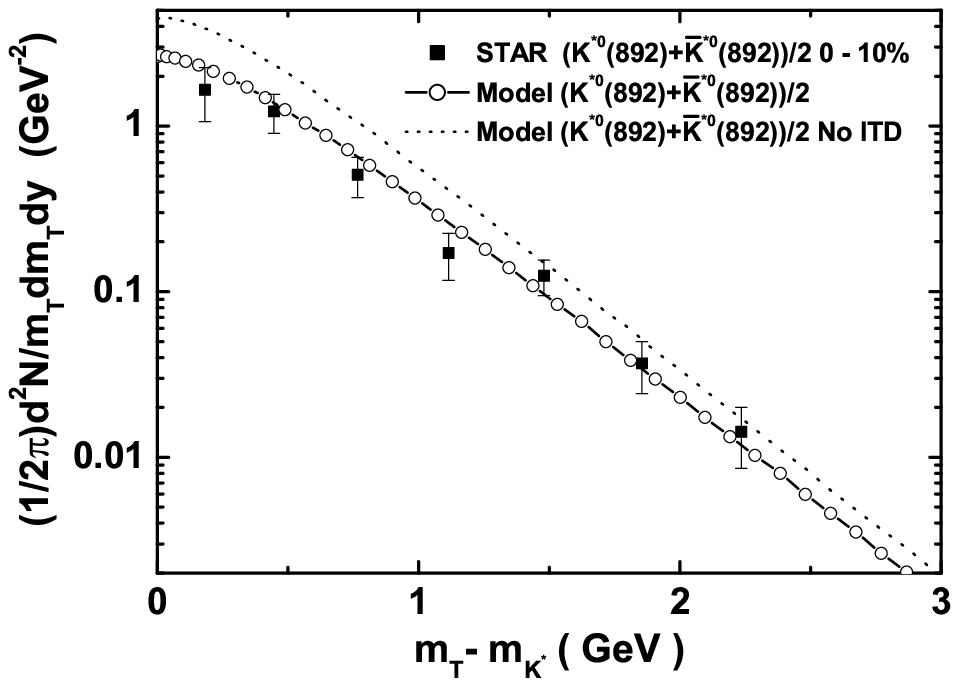}
    \caption{The inclusive $m_T$ distribution of
    $(K^{*0}(892)+\overline{K}^{*0}(892))/2$. The dotted line is
    the model calculation without considering the in-transport
    decay (ITD) loss correction. The data are from
    STAR Collaboration \cite{sign:exp_STAR_K892}.}
    \label{fig13}
\end{figure}

\begin{figure}[h]
 \centering
    \includegraphics[width=0.45\textwidth]{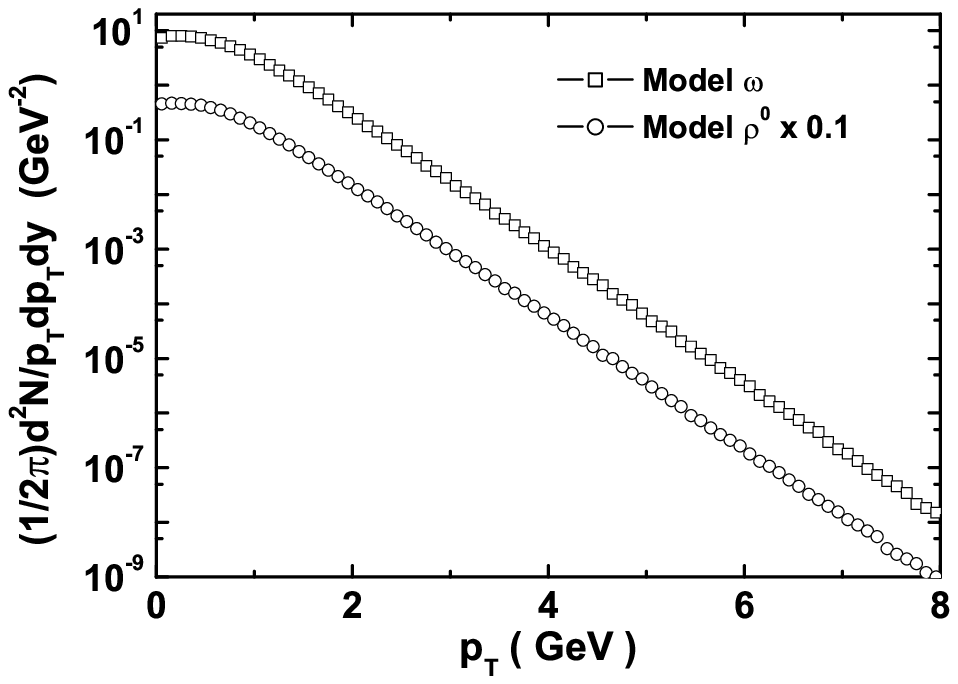}
    \caption{The model predicted inclusive distributions of $\rho^0$ and $\omega$.}
    \label{fig14}
\end{figure}

The $K^*(892)$ yield is shown in Fig. \ref{fig13} as a function of
$m_T-m_{K^*}$, where $m_T=\sqrt{p_T^2+m_{K^*}^2}$ is the particle
transverse mass. From the comparison with the STAR data, the decay
correction from the loss term (\ref{dloss}) plays an important
role. As the decay products may be absorbed by the medium, or
their momenta may be shifted, it is hard to obtain the
reconstruction. Since the meson inelastic cross section is
supposed to be $2/3$ of the baryon cross section, the contribution
from ITD loss for mesons would be more significant than the one
for baryons. The transverse momentum distributions for $\rho^0$
and $\omega$ are predicted in Fig. \ref{fig14}.

\begin{figure}[h]
    \centering
    \includegraphics[width=0.45\textwidth]{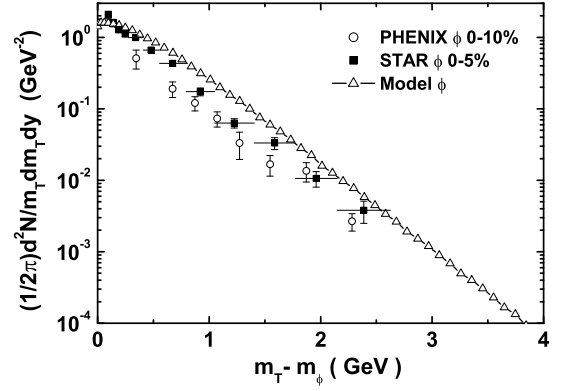}
    \caption{The inclusive $m_T$ distribution of $\phi(1020)$.
    The $x$ error bar is used here to show the $m_T$ bin size.
    The data are from
    STAR \cite{sign:exp_STAR_phi} and PHENIX \cite{sign:exp_PHENIX_phi} Collaborations.}
    \label{fig15}
\end{figure}

The $\phi$ production is shown in Fig.\ref{fig15}. The model
results are much larger than the experimental data. This is
partially due to the limited hadron list in our treatment.
Otherwise, the emission at the hypersurface for $\phi$ may not be
effective. Note the fact that the STAR data
\cite{sign:exp_STAR_phi} are about twice the PHENIX data
\cite{sign:exp_PHENIX_phi}, further precise measurement for $\phi$
is required.

\begin{figure}[h]
    \centering
    \includegraphics[width=0.45\textwidth]{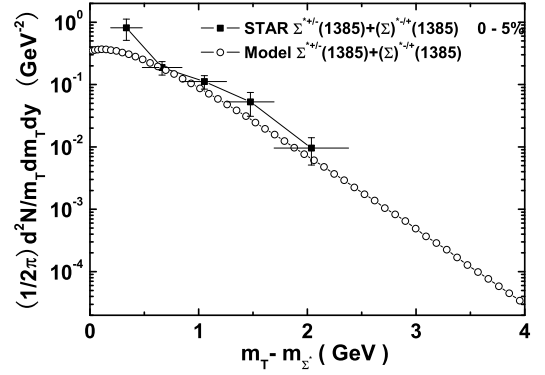}
    \caption{The inclusive $m_T$ distribution of $\Sigma^{*\pm}(1385) + \overline{\Sigma}^{*\mp}(1385)$.
    The data are from STAR Collaboration \cite{sign:exp_STAR_Sig1385}.}
    \label{fig16}
\end{figure}

The $\Sigma^*(1385)$ distribution is shown in Fig. \ref{fig16}.
The model results are much smaller than the experimental data,
especially in very low $p_T$ region. The situation here is similar
to the case for $\Lambda$. The $\Lambda \pi$ or $N\bar K$ process
which is excluded in the model may regenerate some
$\Sigma^*(1385)$
\cite{sign:exp_STAR_Sig1385,sign:exp_STAR_Xi1530}. As
$\Sigma^*(1385)$ can not decay to $N\bar{K}$ due to its low mass,
it may become a one-way valve to deliver strange quarks from $K$
to $\Lambda(\Sigma)$. Another possible reason for the $\Sigma^*$
suppression is that many heavy resonances which would decay to
$\Sigma^*(1385)$ are not included in our calculation.

\begin{figure}[h]
    \centering
    \includegraphics[width=0.45\textwidth]{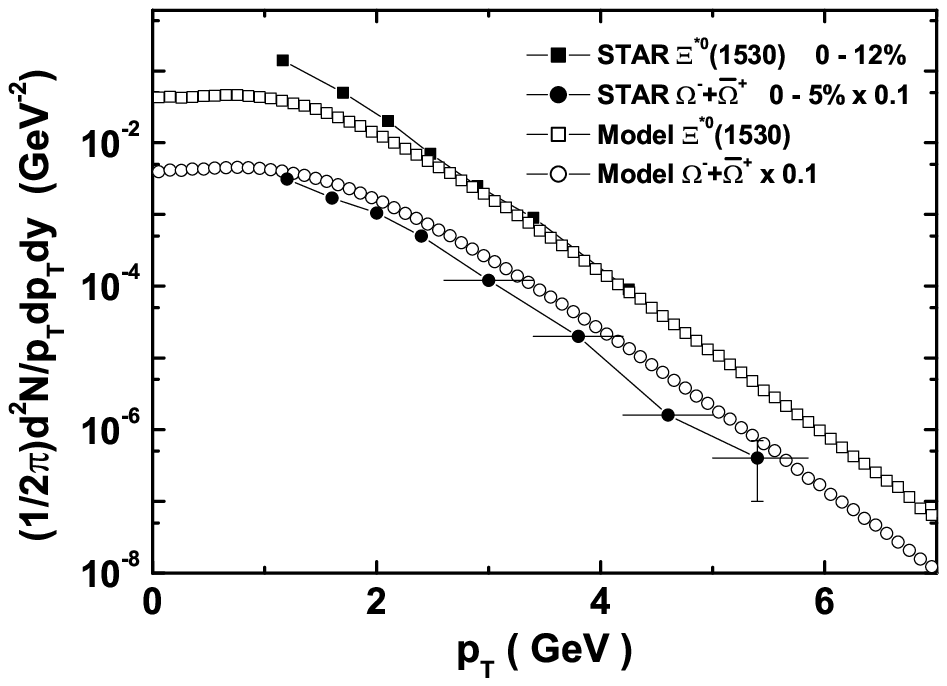}
    \caption{The inclusive transverse momentum distributions of $\Xi^{0}(1530)$
    and $\Omega^-+{\bar{\Omega}}^+$. The data are from STAR Collaboration
    \cite{sign:exp_STAR_Xi1530,sign:exp_STAR_StrangeOld,sign:exp_STAR_Strange}.}
    \label{fig17}
\end{figure}

The transverse momentum distributions of $\Xi^{*0}(1530)$ and
$\Omega$ are shown in Fig.\ref{fig17}. For $\Omega$, the
calculated distribution is close to the data in low $p_T$ region
but deviate from the data in high $p_T$ region. The situation is
similar to $\phi$. This shows that either more resonances should
be considered or the scenario of the hypersurface emission for
these hadrons need to be corrected. From the data, it looks that
there are two slope parameters for $\Omega$, like $\Lambda$ and
$\Sigma^*(1385)$. As for $\Xi^{*0}(1530)$, the model calculation
is bad at low $p_T$. It might be due to the fact that $\Xi(1950)$
is not counted in our model. $\Xi(1950)$ is a mysterious baryon.
Its spin, parity and branch ratios are not clear, and it is even
suggested that there might be more than one $\Xi$ at that mass
\cite{sign:PDG2006}.

\begin{figure}[h]
    \centering
    \includegraphics[width=0.45\textwidth]{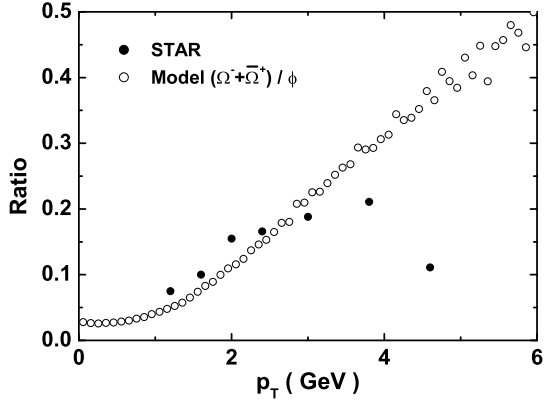}
    \caption{The inclusive ratio $(\Omega^-+{\bar{\Omega}}^+)/\phi$ as a function of transverse momentum.
    The data are from STAR Collaboration \cite{sign:exp_Ratio_Omegaphi}.}
    \label{fig18}
\end{figure}

The ratio of $\Omega$ to $\phi$ is presented in Fig.\ref{fig18} as
a function of transverse momentum.  As multi-strange hadrons, the
jet contribution to $\phi$ and $\Omega$ is expected to be small,
as we have seen in their spectra in Figs.\ref{fig15} and
\ref{fig17}. Our results are similar to the ones calculated in
recombination model \cite{sign:reco_S}.

\begin{figure}[h]
    \centering
    \includegraphics[width=0.45\textwidth]{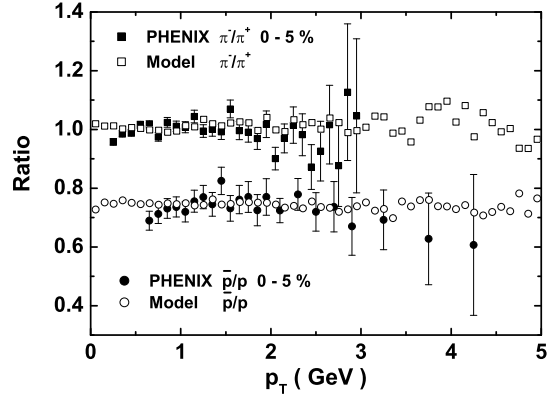}
    \caption{The inclusive $\pi^-/\pi^+$ and exclusive $\bar{p}/p$
    ratios as functions of transverse momentum. The data are from PHENIX Collaboration \cite{sign:exp_PHENIX_ppik}.}
    \label{fig19}
\end{figure}

\begin{figure}[h]
    \centering
    \includegraphics[width=0.45\textwidth]{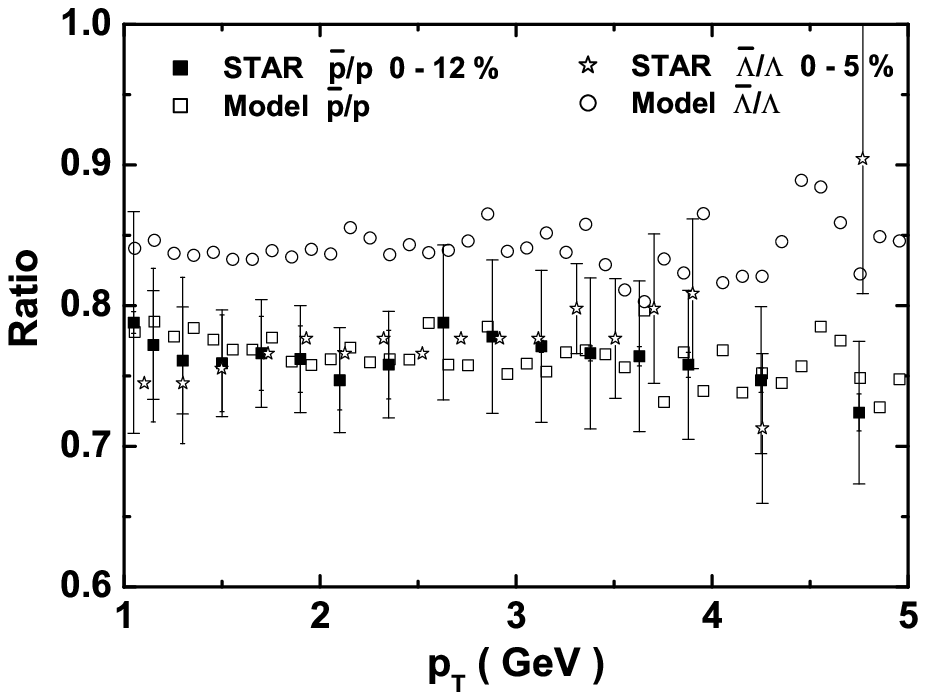}
    \caption{The inclusive $\bar{p}/p$ and $\bar{\Lambda}/\Lambda$
    ratios as functions of transverse momentum. The data are from STAR Collaboration \cite{sign:exp_STAR_ppik2006,sign:exp_STAR_Strange2}.}
    \label{fig20}
\end{figure}

Some of the momentum integrated ratios are listed in Table
\ref{tab:ratio1} and compared with experimental data
\cite{sign:exp_PHENIX_ppik,sign:exp_Ratio_STAR_Omega}. The data
from \cite{sign:exp_Ratio_STAR_Omega} correspond to baryons in the
$p_T$ region of $2.4-3.6$ GeV and mesons in the $p_T$ region of
$1.6-2.4$ GeV. The influence of hadron decay is relatively small
in these regions. The strange baryon ratios at colliding energy
$\sqrt s =200$ GeV \cite{sign:exp_Ratio_STAR_Omega} are very
similar to the data at energy $\sqrt s=130$ GeV \cite{sign:sign
harris,sign:sign Lambda_STAR,sign:sign Lambda_PHENIX}. Some meson
and baryon ratios as functions of transverse momentum are shown in
Figs.\ref{fig19} and \ref{fig20}.
\begin{table}[h]
    \begin{center}
      \begin{tabular}{l c c l}
        \hline\hline $Particles$ & $Model$ & \quad & {$\qquad Exp$} \\
        \hline
            $\pi^-/\pi^+$(in) & 1.006 & & $0.984 \pm 0.004 \pm 0.057$ \cite{sign:exp_PHENIX_ppik}\\
            $\pi^-/\pi^+$(ex) & 0.999 & & $1.01 \pm 0.02$ \cite{sign:exp_Ratio_STAR_Omega}\\
            $K^-/K^+$         & 0.915 & & $0.933 \pm 0.007 \pm 0.054$ \cite{sign:exp_PHENIX_ppik}\\
                              &       & & $0.96 \pm 0.03$ \cite{sign:exp_Ratio_STAR_Omega}\\
            $\bar{p}/p$(ex)   & 0.748 & & $0.731 \pm 0.011 \pm 0.062$ \cite{sign:exp_PHENIX_ppik}\\
            $\bar{p}/p$(in)   & 0.783 & & $0.747 \pm 0.007 \pm 0.046$ \cite{sign:exp_PHENIX_ppik}\\
                              &       & & $0.77 \pm 0.05$ \cite{sign:exp_Ratio_STAR_Omega}\\
            $\overline{{\Lambda}}/\Lambda$
                              & 0.844 & & $0.72 \pm 0.024$ \cite{sign:exp_Ratio_STAR_Omega}\\
            $\overline{\Xi}^+/\Xi^-$
                              & 0.907 & & $0.82 \pm 0.05$ \cite{sign:exp_Ratio_STAR_Omega}\\
            $\overline{\Omega}^+/\Omega^-$
                              & 1.011 & & $1.01 \pm 0.08$ \cite{sign:exp_Ratio_STAR_Omega}\\
            $K^+/\pi^+$       & 0.203 & & $0.171 \pm 0.001 \pm 0.010$ \cite{sign:exp_PHENIX_ppik}\\
            $K^-/\pi^-$       & 0.184 & & $0.162 \pm 0.001 \pm 0.010$ \cite{sign:exp_PHENIX_ppik}\\
            $p_{_{ex}}/\pi^+$ & 0.0645& & $0.064 \pm 0.001 \pm 0.003$ \cite{sign:exp_PHENIX_ppik}\\
            $p/\pi^+$(in)     & 0.102 & & $0.099 \pm 0.001 \pm 0.006$ \cite{sign:exp_PHENIX_ppik}\\
            $\bar{p}_{_{ex}}/\pi^-$     & 0.0479& & $0.047 \pm 0.001 \pm 0.002$ \cite{sign:exp_PHENIX_ppik}\\
            $\bar{p}/\pi^-$(in)         & 0.0797& & $0.075 \pm 0.001 \pm 0.004$ \cite{sign:exp_PHENIX_ppik}\\
        \hline\hline
      \end{tabular}
    \end{center}
 \caption{Some momentum integrated hadron ratios compared with RHIC data \cite{sign:exp_PHENIX_ppik,sign:exp_Ratio_STAR_Omega}.
 The statistical errors in the model calculations are up to $5\%$.}
    \label{tab:ratio1}
\end{table}

$\Sigma^0$ and $\Lambda$ particles are sensitive to the
hadronization mechanisms. In some string fragmentation models the
ratio $\Sigma^0/\Lambda$ is taken to be $\sim 0.35$, but in
thermal statistical models it is predicted to be $0.65-0.75$ due
to the mass difference. If the hadronization is dominated by gluon
junction or coalescence, the ratio is about $1$. In our case, the
hadronization mechanism is similar to the coalescence or
recombination, but the ratio is a little larger than $1$
\cite{sign:diquark_baryon,sign:diquark_Omega}. The reason is that
$\Sigma^0$ contains an important component
$B_\frac{1}{2}(A_{ud},s)$, described in Appendix
\ref{app:wavefunc}, but this term dos not exist for $\Lambda$ with
isospin zero.

\begin{figure}[h]
        \centering
        \includegraphics[width=0.45\textwidth]{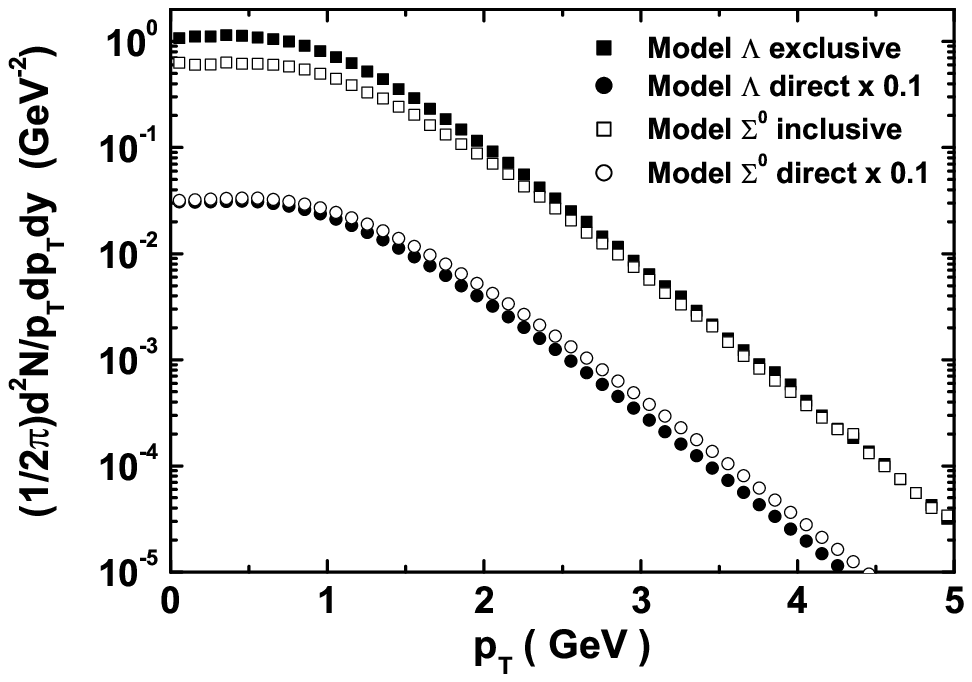}
        \caption{The model predicted transverse momentum distributions of direct $\Sigma^0$ and $\Lambda$,
            inclusive $\Sigma^0$, and exclusive $\Lambda$. The decay contribution from $\Sigma^0$, $\Xi^-$, $\Xi^0$
            and $\Omega$ to the exclusive $\Lambda$ are removed.}
        \label{fig21}
\end{figure}

\begin{figure}[h]
         \centering
        \includegraphics[width=0.45\textwidth]{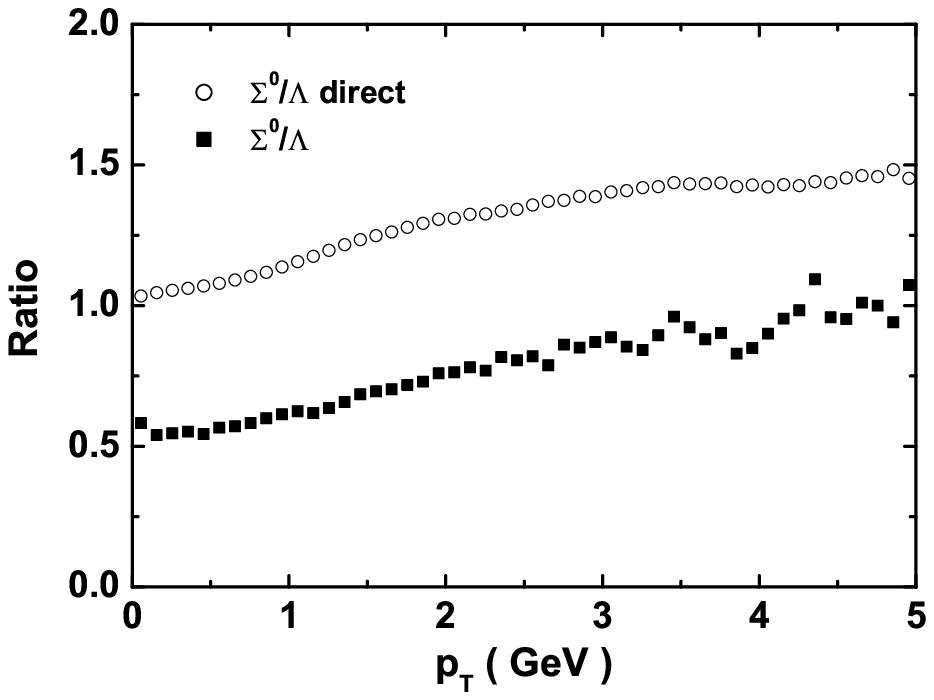}
        \caption{The ratio $\Sigma^0/\Lambda$ as a function of transverse momentum,
        with and without considering decay contribution. }
        \label{fig22}
 \end{figure}

With the selected parameters, the direct ratio $\Sigma^0/\Lambda$
is $1.181$. Considering the decay contributions from heavy baryons
or resonances, it decreases to $0.636$. The transverse momentum
distributions of direct $\Sigma^0$ and $\Lambda$, inclusive
$\Sigma^0$, and exclusive $\Lambda$ without $\Sigma^0$ decay
contribution are shown in Fig.\ref{fig21}. The transverse momentum
dependence of the ratio $\Sigma^0/\Lambda$ with and without
considering the decay contribution are shown in Fig. \ref{fig22}.

\begin{figure}[h]
        \centering
        \includegraphics[width=0.45\textwidth]{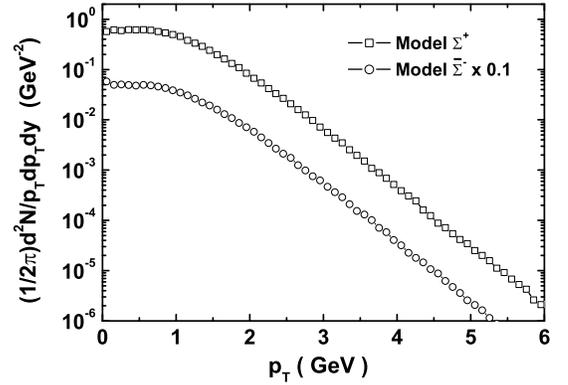}
        \caption{The expected transverse momentum distributions of direct $\Sigma^+$and $\bar{\Sigma}^-$.
        No feed-down correction is applied.}
        \label{fig23}
\end{figure}

Since it is difficult to extract $\Sigma^0$ through
electromagnetic decay, $\Sigma^+$ may be easier to be determined
if $\pi^0$ is well measured. Its production is supposed to be
similar to $\Sigma^0$. The expected $\Sigma^+$ production is shown
in Fig. \ref{fig23}.

\begin{figure}[h]
        \centering
        \includegraphics[width=0.45\textwidth]{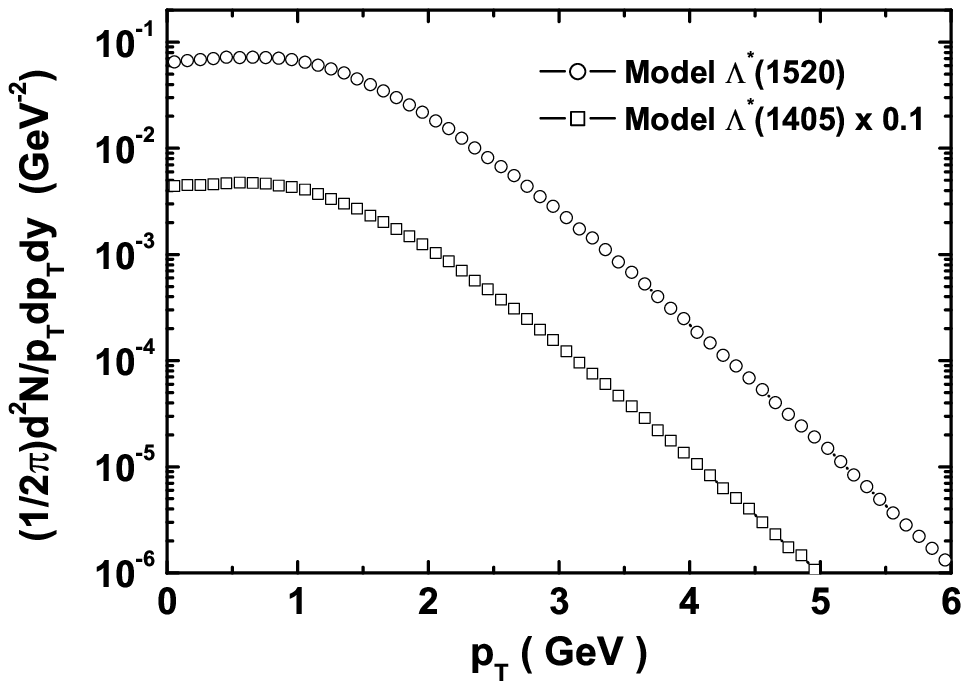}
        \caption{The model calculated transverse momentum distributions of $\Lambda(1405)$ and $\Lambda(1520)$, with the wave functions listed
        in Appendix \ref{app:wavefunc}. No feed-down correction is applied. }
        \label{fig24}
\end{figure}

$\Lambda(1405)$ $0(\frac{1}{2}^-)$ and $\Lambda(1520)$
$0(\frac{3}{2}^-)$ are regarded as singlet states. It is not easy
to determine their wave functions in quark-diquark model, and it
is hard to distinguish them from the states $(L=1, S=\frac{3}{2})$
and $(L=1, S=\frac{1}{2})$. A rough estimation of $\Lambda(1405)$
and $\Lambda(1520)$ production is shown in Fig. \ref{fig24}, where
$\Lambda(1405)$ and $\Lambda(1520)$ are, respectively, supposed to
be $(L=1, S=\frac{1}{2})$ octet and $(L=1, S=\frac{3}{2})$ octet.
The details on their wave function construction is given in
Appendix \ref{app:wavefunc}.

At the end of the discussion on hadron spectra, we consider the
ratio $\bar\Omega^+/\Omega^-$. When diquarks are considered as
constituents of the strongly coupled QGP, the strange chemical
potential of the system will increase due to the negative strange
numbers of diquarks $us$ and $ds$. In our calculations, the
strange chemical potential $\mu_S = 9.35$ MeV is extracted from
the strange charge conservation. In this case, the strange quark
and diauark $ss$ will carry negative chemical potentials, $\mu_s =
-0.35$ MeV and $\mu_{ss} = -0.7$ MeV \cite{sign:diquark_Omega}.
The string fragmentation model \cite{sign:Omega/OmegaString} with
diquarks gives similar result. However, the ratio in the
calculation without diquarks is $\overline{\Omega}^+/\Omega^- \leq
1$ \cite{sign:coal_3,sign:itro_Rafelski_phL}.

As shown in Table \ref{tab:ratio1}, the ratio
$\overline{\Omega}^+/\Omega^- = 1.011$ is consistent with the STAR
measurement \cite{sign:exp_Ratio_STAR_Omega}. While the diquarks
are supposed in thermal equilibrium in our model, and the current
data are still with large statistical and systematic errors, the
qualitative conclusion of $\overline{\Omega}^+/\Omega^-
> 1$ is supposed not to be changed.

As pointed above, the thermalization of final state hadrons are
omitted in all of the calculations. Considering the interactions
which are not inserted in our treatments, some of the dominant
hadrons would be partially thermalized effectively. Comparing with
the RHIC data, $\pi$, $K$, $p$ and $\Lambda$ may be well
thermalized, while the heavy baryons may be not.

\section{Summary}
\label{s5}
We have developed a phenomenological model to describe hadron
production in relativistic heavy ion collisions. In our model, the
QGP evolution and hadron transport are coupled to each other in a
self-consistent way. The strongly coupled QGP, including not only
gluons and quarks but also their resonant states like diquarks, is
described by ideal hydrodynamics, and the hadrons emitted from the
QGP satisfy transport equations with loss and gain terms. The
hadron decay is effectively reflected in the medium absorption and
hadron spectra. Due to the feedback of the hadron emission, the
hadronization hypersurface of the parton fluid shrinks with a
constant velocity. The splitting of the fluid due to its fast
expansion in the later stage and the hadron emission is also
effectively considered, by introducing a ratio of the volume of
the QGP droplets to the whole volume of the system. We applied our
model to calculate the final state hadron spectra in heavy ion
collisions. From the comparison with the RHIC data, most of the
hadron spectra and hadron ratios can be reasonably described in
our model.

The model deserves further improvement in future studies. Unlike
quarks and gluons, diquarks in QGP are resonant states and they
should behave differently. As a consequence, the assumption that
the baryon cross section is $1.5$ times the meson cross section,
which arises from the constituent quark model, should be
reconsidered carefully. Since mesons are described by transport
equations in our model, the assumption of diquark thermalization
is in principle not self-consistent. Different types of hadrons
will have different thermalization time. Light hadrons are easy
but heavy hadrons are difficult to get thermalized. In our current
treatment, the relaxation time in hadron transport equations is
taken to be infinity, and therefore, no hadron can be thermalized
in the whole evolution. The treatment of the QGP fluid splitting
is also too simple in our model. When the model is improved, it
can be used to calculate more hadron distributions like elliptic
flow\cite{sign:v2flow01,sign:phiflow}, which is considered to be
sensitive to the thermalization of the parton system in the early
stage of heavy ion collisions.\vspace{12pt}
\\
{\bf Acknowledgements:} We are grateful to Wojciech Broniowski,
Huanzhong Huang, Rudolph C. Hwa, Yuxin Liu, Sevil Salur, Qubing
Xie, Nu Xu and Weining Zhang for useful discussions. The work is
supported in part by the grants No. NSFC10547001 and 10425810.

\appendix
\section{Rewriting the Baryon Wave Functions}
\label{app:wavefunc}
The SU(6) baryon wave functions can be
rewritten in the quark-diquark model
\cite{sign:diquark_form,sign:diquark_Lamda}. The $(\frac{1}{2})^+$
and $(\frac{3}{2})^+$ s-wave intermediate baryon states are
introduced as
\begin{eqnarray}
B_{\frac{1}{2},+{1\over 2}}(A,q)&=&{1\over \sqrt 3}\left[\sqrt 2 A_{+}q^{\downarrow}-A_0q^{\uparrow}\right],\nonumber\\
B_{\frac{1}{2},+{1\over 2}}(S,q)&=&Sq^{\uparrow},\nonumber\\
B_{\frac{3}{2},+{3\over 2}}(A,q)&=&A_{+}q^{\uparrow},\nonumber\\
B_{\frac{3}{2},+{1\over 2}}(A,q)&=&{1\over \sqrt
3}\left[A_{+}q^{\downarrow}+\sqrt 2 A_0q^{\uparrow}\right],
\end{eqnarray}
where $S$ and $A$ represent a scalar and an axial-vector diquark
state, respectively. The final baryon states can be expressed in
terms of certain combinations of these intermediate baryon states
\cite{sign:diquark_baryon,sign:diquark_Omega}. The wave functions
are expressed as
\begin{eqnarray}
 |{p}\rangle&=&\frac{1}{\sqrt{3}}\Big[B_\frac{1}{2}(A_{uu},d)-\sqrt{\frac{1}{2}}B_\frac{1}{2}(A_{ud},u)\nonumber\\
 && +\sqrt{\frac{3}{2}}B_\frac{1}{2}(S_{ud},u)\Big],\nonumber\\
 |{n}\rangle&=&\frac{1}{\sqrt{3}}\Big[-B_\frac{1}{2}(A_{dd},u)+\sqrt{\frac{1}{2}}B_\frac{1}{2}(A_{ud},d)\nonumber\\
 && +\sqrt{\frac{3}{2}}B_\frac{1}{2}(S_{ud},d)\Big],\nonumber\\
 |{\Lambda}\rangle&=&\frac{1}{\sqrt{3}}\Big[\sqrt{\frac{3}{4}}B_\frac{1}{2}(A_{us},d)-\sqrt{\frac{3}{4}}B_\frac{1}{2}(A_{ds},u)\nonumber\\
 &&+\sqrt{\frac{1}{4}}B_\frac{1}{2}(S_{us},d)-\sqrt{\frac{1}{4}}B_\frac{1}{2}(S_{ds},u)\nonumber\\
 &&+B_\frac{1}{2}(S_{ud},s)\Big],\nonumber\\
 |{\Sigma^0}\rangle&=&\frac{1}{\sqrt{3}}\Big[B_\frac{1}{2}(A_{ud},s)-\sqrt{\frac{1}{4}}B_\frac{1}{2}(A_{us},d)\nonumber\\
 &&-\sqrt{\frac{1}{4}}B_\frac{1}{2}(A_{ds},u)+\sqrt{\frac{3}{4}}B_\frac{1}{2}(S_{us},d)\nonumber\\
 &&+\sqrt{\frac{3}{4}}B_\frac{1}{2}(S_{ds},u)\Big],\nonumber\\
 |{\Sigma^+}\rangle&=&-\frac{1}{\sqrt{3}}\Big[B_\frac{1}{2}(A_{uu},s)-\sqrt{\frac{1}{2}}B_\frac{1}{2}(A_{us},u)\nonumber\\
 &&+\sqrt{\frac{3}{2}}B_\frac{1}{2}(S_{us},u)\Big],\nonumber\\
 |{\Sigma^-}\rangle&=&\frac{1}{\sqrt{3}}\Big[B_\frac{1}{2}(A_{dd},s)-\sqrt{\frac{1}{2}}B_\frac{1}{2}(A_{ds},d)\nonumber\\
 &&+\sqrt{\frac{3}{2}}B_\frac{1}{2}(S_{ds},d)\Big],\nonumber\\
 |{\Xi^0}\rangle&=&\frac{1}{\sqrt{3}}\Big[-B_\frac{1}{2}(A_{ss},u)+\sqrt{\frac{1}{2}}B_\frac{1}{2}(A_{us},s)\nonumber\\
 &&+\sqrt{\frac{3}{2}}B_\frac{1}{2}(S_{us},s)\Big],\nonumber\\
 |{\Xi^-}\rangle&=&\frac{1}{\sqrt{3}}\Big[-B_\frac{1}{2}(A_{ss},d)+\sqrt{\frac{1}{2}}B_\frac{1}{2}(A_{ds},s)\nonumber\\
 && +\sqrt{\frac{3}{2}}B_\frac{1}{2}(S_{ds},s)\Big]
\end{eqnarray}
for the SU(3) octet baryons, and
\begin{eqnarray}
 |{\Delta^{++}}\rangle &=&B_\frac{3}{2}(A_{uu},u),\nonumber\\
 |{\Delta^+}\rangle&=&\frac{1}{\sqrt{3}}\Big[B_\frac{3}{2}(A_{uu},d)+\sqrt{2}B_\frac{3}{2}(A_{ud},u)\Big],\nonumber\\
 |{\Delta^0}\rangle&=&\frac{1}{\sqrt{3}}\Big[B_\frac{3}{2}(A_{dd},u)+\sqrt{2}B_\frac{3}{2}(A_{ud},d)\Big],\nonumber\\
 |{\Delta^{-}}\rangle &=&B_\frac{3}{2}(A_{dd},d),\nonumber\\
 |{\Sigma^{*+}}\rangle&=&\frac{1}{\sqrt{3}}\Big[B_\frac{3}{2}(A_{uu},s)+\sqrt{2}B_\frac{3}{2}(A_{us},u)\Big],\nonumber\\
 |{\Sigma^{*0}}\rangle&=&\frac{1}{\sqrt{3}}\Big[B_\frac{3}{2}(A_{us},d)+B_\frac{3}{2}(A_{ds},u)+B_\frac{3}{2}(A_{ud},s)\Big],\nonumber\\
 |{\Sigma^{*-}}\rangle&=&\frac{1}{\sqrt{3}}\Big[B_\frac{3}{2}(A_{dd},s)+\sqrt{2}B_\frac{3}{2}(A_{ds},d)\Big],\nonumber\\
 |{\Xi^{*0}}\rangle&=&\frac{1}{\sqrt{3}}\Big[B_\frac{3}{2}(A_{ss},u)+\sqrt{2}B_\frac{3}{2}(A_{us},s)\Big],\nonumber\\
 |{\Xi^{*-}}\rangle&=&\frac{1}{\sqrt{3}}\Big[B_\frac{3}{2}(A_{ss},d)+\sqrt{2}B_\frac{3}{2}(A_{ds},s)\Big],\nonumber\\
 |{\Omega^{-}}\rangle &=&B_\frac{3}{2}(A_{ss},s)
\end{eqnarray}
for the decuplet baryons. From these expressions, we can get the
factor $C_I$ in Eqs. (\ref{ci1}) and (\ref{ci2}) to complete the
production rate.

There might be other orthogonal baryon states. From the Tables of
Particle Physics \cite{sign:PDG2006}, $N(1440)$
$\frac{1}{2}(\frac{1}{2}^+)$ and $N(1710)$
$\frac{1}{2}(\frac{1}{2}^+)$ are conjectured to be the orthogonal
states of $p$ and $n$. Other possible orthogonal states are
$\Lambda(1600)$ $0(\frac{1}{2}^+)$, $\Lambda(1810)$
$0(\frac{1}{2}^+)$, $\Sigma(1660)$ $1(\frac{1}{2}^+)$ and
$\Sigma(1880)$ $1(\frac{1}{2}^+)$ of $\Lambda$ and $\Sigma$,
$N(1720)$ $\frac{1}{2}(\frac{3}{2}^+)$ of $\Delta^+$ and
$\Delta^0$, $\Lambda(1890)$ $0(\frac{3}{2}^+)$ and $\Sigma(1840)$
$1(\frac{3}{2}^+)$ of $\Sigma^*(1385)$, and some unknown
orthogonal states of $\Xi$ and $\Xi^*(1530)$. There are 24 such
particles which belong to two $(\frac{1}{2})^+$ octets and a
$(\frac{3}{2})^+$ octet.

The wave functions of the orthogonal baryon states in
$(\frac{3}{2})^+$ decuplet can be determined. For example,
\begin{eqnarray}
|{\Lambda}_2\rangle^{\{8\}}_{S=\frac{3}{2}} &\!\!=&\!\!
\frac{1}{\sqrt{2}}\left[B_\frac{3}{2}(A_{us},d)-B_\frac{3}{2}(A_{ds},u)\right],\\
|{\Sigma^{0}_2}\rangle^{\{8\}}_{S=\frac{3}{2}} &\!\!=&\!\!
\frac{1}{\sqrt{6}}\left[B_\frac{3}{2}(A_{us},d)+B_\frac{3}{2}(A_{ds},u)-2B_\frac{3}{2}(A_{ud},s)\right].\nonumber
\end{eqnarray}

There are two orthogonal baryon states in $(\frac{1}{2})^+$ octet.
Their wave functions are not unique. Considering that the first
octet state is the mixed state of $Aq$ and $Sq$, a most possible
assumption is that the second octet state should be the
corresponding mixed state, and the third one contains no mixing.
Therefore, we have
\begin{eqnarray}
|{N^+_2}\rangle^{\{8\}}_{S=\frac{1}{2}} &\!\!=&\!\!
    \frac{1}{\sqrt{3}}\Big[B_\frac{1}{2}(A_{uu},d)-\sqrt{\frac{1}{2}}B_\frac{1}{2}(A_{ud},u)\\
    &&-\sqrt{\frac{3}{2}}B_\frac{1}{2}(S_{ud},u)\Big],\nonumber\\
 |{N^+_3}\rangle^{\{8\}}_{S=\frac{1}{2}} &\!\!=&\!\!
    \frac{1}{\sqrt{3}}\left[B_\frac{1}{2}(A_{uu},d)+\sqrt{2}B_\frac{1}{2}(A_{ud},u)\right],\nonumber
\end{eqnarray}
where, $N_2^+$ corresponds to $N(1440)^+$ and $N_3^+$ to
$N(1710)^+$. The other orthogonal baryon wave functions with
positive parity are similar.

The wave functions of some of the $(\frac{1}{2})^-$ and
$(\frac{3}{2})^-$ resonances are not clear. It is hard to
distinguish a $(\frac{1}{2})^-$ or a $(\frac{3}{2})^-$ state from
the two candidates $(L=1, S=\frac{1}{2})$ and $(L=1,
S=\frac{3}{2})$. For simplicity, all $(\frac{1}{2})^-$ baryons,
except $\Delta$, are regarded as $S=\frac{1}{2}$ octet states and
all $(\frac{3}{2})^-$ baryons are regarded as $S=\frac{3}{2}$
octet or decuplet states.

$\Lambda(1405)$ $0(\frac{1}{2}^-)$ and $\Lambda(1520)$
$0(\frac{3}{2}^-)$ are regarded as singlet states
\cite{sign:PDG2006}. The quark-diquark wave functions of them are
unknown. Approximately, Their wave functions are written as
$S=\frac{1}{2}$ octet and $S=\frac{3}{2}$ octet.

\section{Effective Lagrangian and Matrix Elements of Production Cross Section}
\label{app:M2}

In calculating the hadron production rate, only lowest order
interactions are considered. As there are not enough data to
determine the coupling modes and constants, we choose only some of
the most simple coupling channels. To further reduce the number of
parameters, we use two universal coupling constants $g_M$ for
meson interactions with two quarks and $g_B$ for baryon
interactions with a quark and a diquark. The meson sector of the
effective Lagrangian density is listed in
Tab.\ref{tab:Lagrangian01} for different meson. The influence of
P-wave and D-wave is neglected.
\begin{table}[h]
    \begin{center}
      \begin{tabular}{cl l }
        \hline\hline
        $Meson$ & $ \quad L_{eff}$ & $ \qquad \qquad |M_{spin}|^2$ \\ \hline
        ${{0}^-}$ &\!\!\!$ig_{_M}\bar{q}_1\gamma_5 q_2 \phi$
                                        & $\frac{g_{M}^2}{2}\left[m_{M}^2-(m_{q1}-m_{q2})^2\right]$ \\
        ${{0}^+}$ &\!\!\!$ig_{_M}\bar{q}_1 q_2 \phi$
                                        & $\frac{g_{M}^2}{2}\left[m_{M}^2-(m_{q1}+m_{q2})^2\right]$\\
        ${{1}^-}$ &\!\!\!$ig_{_M}\bar{q}_1\gamma_\mu q_2 \omega_\mu$
                                        & $\frac{g_{M}^2\left[m_M^2-(m_{q1}-m_{q2})^2\right]}{2}\left[\frac{(m_{q1}+m_{q2})^2}{m_M^2}+2\right]$ \\
        ${{1}^+}$ &\!\!\!$ig_{_M}\bar{q}_1 \gamma_\mu\gamma_5 q_2 \omega_\mu$
                                        & $\frac{g_{M}^2\left[m_M^2-(m_{q1}+m_{q2})^2\right]}{2}\left[\frac{(m_{q1}-m_{q2})^2}{m_M^2}+2\right]$\\
        ${{2}^+}$ &\!\!\!\!\!\!$ig_{_M}\bar{q}_1\gamma_\mu \gamma_\nu q_2 \kappa_{\mu\nu}$
                                        & $\frac{3g_{M}^2}{10}\left[m_M^2-(m_{q1}+m_{q2})^2\right]$ \\
        \hline\hline
      \end{tabular}
    \end{center}
 \caption{The effective interactions and the corresponding scattering matrix
 elements $|M_{spin}|^2$ = $|M|^2/3$
 for different mesons. }
 \label{tab:Lagrangian01}
\end{table}

For the baryon sector of the effective Lagrangian density, we take
\begin{eqnarray}
 L_{qSB} &=& ig_{B}\;\bar{B} q S,\\
 L_{qAB1}&=& ig_{B}\;\bar{B}\gamma_{\mu}\gamma_{5}q
 A_{\mu},\nonumber\\
 |M|^2_S &\!\!=&\!\!
    {g_{B}^2}\left[(m_B+m_q)^2-m_{S}^2\right],\nonumber\\
 |M|^2_A &\!\!=&\!\!
    \frac{g_{B}^2}{3}\left[(m_B+m_{q})^2-m_A^2\right]\left[\frac{(m_B-m_{q})^2}{m_A^2}+2\;\right]\nonumber
\end{eqnarray}
for the baryons with $J^P=\frac{1}{2}^+$, $L=0$ and
$S=\frac{1}{2}$. Note that the scattering matrix elements here are
equivalent to the expressions in \cite{sign:diquark_Omega}.

For $J^P=\frac{1}{2}^-$ \cite{sign:Lag_123}, $L=1$ and
$S=\frac{1}{2}$, we use
\begin{eqnarray}
 L_{qSB} &=& \frac{ig_{B}}{m_B+m_q}\;\bar{B} \gamma_5 \gamma_\mu q \partial_\mu S,\\
 L_{qAB1}&=& \frac{ig_{B}}{2m_A}\;\bar{B} \sigma_{\mu\nu} q \partial_\nu
 A_\mu,\nonumber\\
 |M|^2_S &\!\!=&\!\!{g_{B}^2}\left[(m_B-m_{q})^2-m_{S}^2\right],\nonumber\\
 |M|^2_A &\!\!=&\!\!
 \frac{g_{B}^2}{6}\left[(m_B-m_{q})^2-m_A^2\right]\left[\frac{(m_B+m_{q})^2}{m_A^2}+\frac{1}{2}\;\right]\nonumber
\end{eqnarray}
with the definition $\sigma_{\mu\nu} =
-i[\gamma_\mu,\gamma_\nu]/2$. The expression of $L_{qSB}$ here
gives the same result as the direct coupling
\cite{sign:Lag1p2min}. For $J^P=\frac{1}{2}^-$, $L=1$ and
$S=\frac{3}{2}$, the scattering matrix elements are supposed to be
the same.

We choose
\begin{eqnarray}
 L_{qAB} &=& ig_{B}\;\bar{B}_\mu\gamma_{5}q A_{\mu},\nonumber\\
|M|^2_A &=&
\frac{g_{B}^2}{18}\left[(m_B-m_q)^2-m_A^2\;\right]\nonumber\\
&&\times\left[\frac{(m_B^2+m_A^2-m_q^2)^2}{m_B^2m_A^2}+8\;\right]
\end{eqnarray}
for $J^P=\frac{3}{2}^+$ \cite{sign:Lag_RS}, $L=0$ and
$S=\frac{3}{2}$, and
\begin{eqnarray}
 L_{qSB} &=& \frac{ig_{B}}{m_S}\;\bar{B}_{\mu} \gamma_5 q \partial_\mu
 S,\nonumber\\
 |M|^2_S &=& \frac{g_{B}^2}{6}\left[(m_B-m_q)^2-m_S^2\right]\nonumber\\
 && \times\left[\frac{(m_B^2+m_S^2-m_q^2)^2}{m_B^2m_S^2}-4\;\right]
\end{eqnarray}
for $J^P=\frac{3}{2}^-$, $L=1$ and $S=\frac{1}{2}$. For the
axial-vector couplings, the matrix element is replaced by
\begin{eqnarray}
 |M|^2_A &=& \frac{g_{B}^2}{18}\left[(m_B+m_q)^2-m_A^2\;\right]\nonumber\\
 &&
 \times\left[\frac{(m_B^2+m_A^2-m_q^2)^2}{m_B^2m_A^2}+8\;\right],
\end{eqnarray}
and the expression for $J^P=\frac{3}{2}^-$, $L=1$ and
$S=\frac{3}{2}$ is the same.

We employ
\begin{eqnarray}
 L_{qAB} &=& \frac{ig_{B}}{m_A}\;\bar{B}_{\mu\nu} q \partial_\mu
 A_{\nu},\nonumber\\
 M|^2_A &=& \frac{g_{B}^2}{48}\left[(m_B+m_{q})^2-m_A^2\right]\nonumber\\
 &&\times\left[\frac{(m_B^2+m_A^2-m_{q}^2)^2}{m_B^2m_A^2}-4\right]\nonumber\\
 &&\times\left[\frac{(m_B^2+m_A^2-m_{q}^2)^2}{m_B^2m_A^2}+5.6\right]
\end{eqnarray}
for $J^P=\frac{5}{2}^-$, $L=1$ and $S=\frac{3}{2}$, where the
expression of the spin-$\frac{5}{2}$ projector can be found in
\cite{sign:Lag_25}, and
\begin{eqnarray}
 L_{qSB} &=& \frac{ig_{B}}{m_S^2}\;\bar{B}_{\mu\nu} q \partial_\mu
 \partial_{\nu}S,\nonumber\\
|M|^2_S &=&\frac{g_{B}^2}{16}\left[(m_B+m_q)^2-m_S^2\right]\nonumber\\
&&
\times\left[\frac{(m_B^2+m_S^2-m_q^2)^2}{m_B^2m_S^2}-4\;\right]^2
\end{eqnarray}
for $J^P=\frac{5}{2}^+$, $L=2$ and $S=\frac{1}{2}$.


\end{document}